\documentclass[a4paper, 10pt]{article}
\usepackage{jheppub} % for details on the use of the package, please
\usepackage{amsmath}
\usepackage{amssymb}
\usepackage{mathrsfs}
\usepackage{physics}
\usepackage{graphicx}
\usepackage[dvipsnames]{xcolor}
\usepackage{subfig} 
\usepackage{empheq}
\usepackage{esint}
\usepackage{hyperref}
\usepackage[shortlabels]{enumitem}
\usepackage{tikz, pgf} 
\usetikzlibrary{arrows}
\usepackage{pgfplots}
\usepackage{natbib} 
\usepackage{float}
\usepackage[mathscr]{euscript}
\restylefloat{table}
\usepackage{cleveref}
\usepackage[bbgreekl]{mathbbol}

\usetikzlibrary{cd}		% commutative diagrams
\usetikzlibrary{calc}		% Calculations in tikz
\usetikzlibrary{math} %needed tikz library for setting variables
\usetikzlibrary{intersections}
\usetikzlibrary{decorations.pathmorphing}
\usetikzlibrary{decorations.pathreplacing}
\usetikzlibrary{decorations.markings,decorations.text}

\tikzset{snake it/.style={decorate, decoration=snake}}
\tikzset{snake light/.style={decorate, decoration={snake, segment length=1.7 mm, amplitude=0.3 mm}}}
\tikzset{->-/.style={decoration={
			markings,
			mark=at position .5 with {\arrow{>}}},postaction={decorate}}}
\tikzset{-<-/.style={decoration={
			markings,
			mark=at position .5 with {\arrow{<}}},postaction={decorate}}}

\crefname{section}{\S\!\!}{\S\S\!\!}
\Crefname{section}{\S}{\S\S}
\crefname{appendix}{Appendix\!}{Appendices\!}
\crefname{figure}{Fig.\!}{Figs.\!}
\crefname{table}{Table\!}{Tables\!}
\crefname{equation}{}{Eqs.\!}
\newcommand{\w}{\omega}

\newcommand{\bD}{\mathbb{D}}

\newcommand{\del}{\partial}
\newcommand{\sig}{\theta}
\newcommand{\csig}{\varsigma}
\newcommand{\bsig}{\sigma}
\newcommand{\isig}{\mathbb{s}}
\newcommand{\bbsig}{\bbsigma}
\newcommand{\vsig}{\vartheta}
\newcommand{\msig}{\mathsf{s}}
\newcommand{\gorb}{\texttt{g}^{(\sig)}}
\newcommand{\flow}{\texttt{F}}
\newcommand{\bx}{\vb{x}}
\newcommand{\bk}{\vb{k}}
\newcommand{\idim}{\mathsf{p}}

\newcommand{\ctor}{\zeta}

\newcommand{\av}{\mathsf{a}}
\newcommand{\dif}{\mathsf{d}}
\newcommand{\sL}{{}{\mathsf{L}}}
\newcommand{\sR}{{}{\mathsf{R}}}
\newcommand{\sSK}{{}_\text{SK}}
\newcommand{\nb}{n_{{}_\text{B}}}

\newcommand{\totr}[1]{\widehat{#1}}

\begin{document}
\title{Real Time Correlations and Complexified Horizons}
\author{Akhil Sivakumar}
\emailAdd{akhil.sivakumar@apctp.org}
\affiliation{Asia Pacific Center for Theoretical Physics\\
	77 Cheongam-ro, Nam-gu, Pohang-si, Gyeongsangbuk-do, 37673, Korea}
\abstract{We construct black hole saddles dual to real-time/Schwinger-Keldysh (SK) path integrals with arbitrary splits of the thermal density matrix generalizing the holographic SK prescription in \cite{Glorioso:2018mmw}. Using a scalar probe on the AdS Schwarzschild black brane as an example, we demonstrate how KMS properties of the boundary correlators naturally derive from these geometries. As deforming the boundary time contour is equivalent to the action of half sided modular transformation, these saddles can be used to compute higher point modular transformed correlators using a well controlled bulk perturbation theory. An interesting relation between these saddles and the more familiar eternal geometries, and the respective generators of time translation on them is described. Inspired from recent discussions on algebras of observables in gravity, we motivate that classical ensembles of such geometries with different amount of modular transformations should be promoted to genuine configurations of the bulk geometry. In particular, this is argued to imply the existence of additional classical moduli in the boundary open EFT dual to the exterior dynamics, and via fluid gravity correspondence, in the fluctuating hydrodynamics of the boundary. Finally, a Lorentzian version of the Kontsevich-Segal conditions is verified for all these geometries. }

\maketitle

%\tableofcontents

\section{Introduction}\label{sec:Intro}

Real time techniques for quantum field theories were originally developed to study near equilibrium response theory \cite{Feynman:1963fq, Schwinger:1960qe, Keldysh:1964ud, Bakshi:1962dv,kamenev_2011}, unitarity proofs for scattering amplitudes \cite{tHooft:1973wag} and more recently primordial perturbations in cosmology \cite{Weinberg:2005vy}.\footnote{See \cite{Chen:2017ryl, DiPietro:2021sjt, Donath:2024utn,Salcedo:2024smn} for some recent discussions in cosmological perturbation theory.} This technology which is commonly called the Schwinger-Keldysh (SK) formalism was introduced into holography in  \cite{Herzog:2002pc} and further clarified in \cite{Skenderis:2008dg}. Its importance in holography basically stems from the observation that effective semi-classical dynamics around black holes is equivalent to the near-equilibrium dynamics of the boundary CFT. Therefore an understanding of this physics in the bulk requires a clean formalism to compute real time correlation functions from gravity. In recent years a highly convenient prescription for the same, originally proposed in \cite{Glorioso:2018mmw} (henceforth called the CGL prescription), has become popular in the literature. Its advantage compared to previous treatments which relied on cut and glue constructions of the relevant gravitational saddle is essentially that, the CGL prescription defines the bulk geometry as a smooth analytical continuation of the black hole solution, which makes it optimal for computing higher order correlation functions using semi-classical perturbation theory. Following \cite{Jana:2020vyx} we will address this geometry as the gravitational Schwinger-Keldysh (grSK) geometry. The CGL prescription has been used to study real time dynamics in various holographic models \cite{Chakrabarty:2019aeu,Loganayagam:2020eue,Ghosh:2020lel,He:2022jnc,Bu:2020jfo,Bu:2021clf,Bu:2021jlp,Bu:2022esd,Banerjee:2022aub,Martin:2024mdm}. Its generalization to charged and rotating backgrounds were investigated in \cite{Loganayagam:2020iol, He:2022deg,Ghosh:2022fyo,Bu:2022oty,Rangamani:2023mok} and \cite{Chakrabarty:2020ohe} respectively (see \cite{Loganayagam:2023pfb} for an adaptation to de Sitter space).  For a discussion on the solution space of field equations and analytical properties of higher point correlation functions see \cite{Loganayagam:2022zmq,Loganayagam:2022teq}.

It is important to distinguish between the real time black hole of the CGL prescription and the more familiar Kruskal extension of the black hole.  Both constructions extend the AdS Schwarzschild solution to a geometry with two asymptotic AdS boundaries. However, they describe different but closely related observables of the CFT. The Kruskal extension is dual to an entangled thermo field double (TFD) state of a pair of identical copies of the CFT given by
\begin{equation}
| \Psi \rangle_{_\text{TFD}} = \mathcal{Z}_{\beta}^{-\frac{1}{2}}\sum_{n}  e^{-\frac{1}{2}\beta E_{n}} | E_{n} \rangle_{_{\sR}} | E_{n} \rangle_{_{\sL}}  \ ,
\end{equation}
where $\mathcal{Z}_{\beta}$ is the thermal partition function and $|E_{n}\rangle_{_{\sR/\sL}}$ are energy eigen states \cite{Maldacena:2001kr}. The $\sR$ and $\sL$ systems here are taken to have opposite directions of time evolution. In the language of SK formalism, corresponding boundary observables  are real time correlations functions of the form 
\begin{equation}\label{eq:TFDObs}
\text{TFD Observables :    }  \qquad \text{tr} \left\{ \left(\rho_{_{\beta}}\right)^{\frac{1}{2}} \mathbb{T} \left( (\mathsf{O}_{1})_{_{\sR}} (\mathsf{O}_{2})_{_{\sR}} \ldots \right) \left(\rho_{_{\beta}}\right)^{\frac{1}{2}} \widetilde{\mathbb{T}} \left( (\mathsf{O}_{1})_{_{\sL}} (\mathsf{O}_{2})_{_{\sL}} \ldots \right) \right\} \ ,
\end{equation}
where $\rho_{_{\beta}} = \mathcal{Z}_{\beta}^{-1}  e^{-\beta \mathsf{H}}$ is the thermal density operator for a single copy of the CFT  and $\mathbb{T} (\widetilde{\mathbb{T}})$ is the (anti) time ordering symbol. They describe correlation functions of two independent strings of time ordered and anti time ordered operator insertions which act on the $\sR$ and $\sL$ copies of the CFT Hilbert space respectively. On the other hand, the CGL prescription yields the following boundary correlation functions which we term the ingoing SK observables.
\begin{equation}\label{eq:SKObs}
\text{Ingoing SK Observables :    }  \qquad\text{tr} \left\{  \mathbb{T} \left( (\mathsf{O}_{1})_{_{\sR}} (\mathsf{O}_{2})_{_{\sR}} \ldots \right) \rho_{_{\beta}} \widetilde{\mathbb{T}} \left( (\mathsf{O}_{1})_{_{\sL}} (\mathsf{O}_{2})_{_{\sL}} \ldots \right) \right\} \ .
\end{equation}

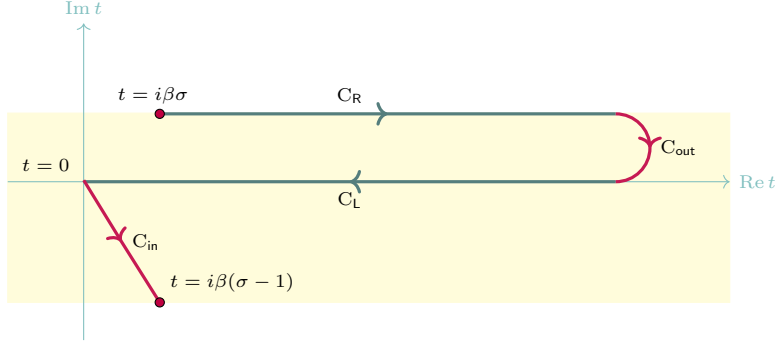
\begin{figure}[h!] 
	\begin{center}
		\begin{tikzpicture}
			
			 \draw[fill,color=yellow!25, opacity=0.7] 	(-1,0.01) -- (8.5,0.01) -- (8.5,-2.5) -- (-1,-2.5) -- cycle; 
			
			\draw[thin, color=teal!50, -> ] (-0.001, -3) -- (-0.001,1.2) node[above]{$\scriptstyle {\Im t}$};
			\draw[thin, color=teal!50, -> ] (-1,-0.9) -- (8.5,-0.9) node[right]{$\scriptstyle {\Re t}$};
			
			\draw[very thick, olive!20!teal,->-] (1,0) node[above]{\scriptsize \textcolor{black}{$t=i \beta \bsig \; $ }} -- (3.5,0) node[above]{\scriptsize \textcolor{black}{C$_{\sR}$}}--(7,0);
			
			\draw[very thick,olive!20!teal,->-] (7,-0.9) -- (3.5,-0.9)node[below]{\scriptsize \textcolor{black}{C$_{\sL}$}}--(0,-0.9)node[above left]{\scriptsize \textcolor{black}{$t=0\, $}};

			\draw[very thick, purple!90, ->-] (-0,-0.88) --(0.5,-1.7) node[right]{\scriptsize \textcolor{black}{ C$_{ \textsf{in}}$}} --(1,-2.5)node[above right]{\scriptsize \textcolor{black}{$t=i\beta (\bsig-1) \, $}};
			
			\draw[very thick, purple!90,->-] (7,0) arc (90: 0 : 0.45)node[right]{\scriptsize \textcolor{black}{ C$_{ \textsf{out}}$}} arc (0: -90 : 0.45);
			\draw[thin,   color=black, fill=purple] (1,0) circle (0.4 ex);
			\draw[thin,  color=black, fill=purple] (1,-2.5) circle (0.4 ex);

		\end{tikzpicture}
	\caption{For the thermal state $\rho_{_\beta} = e^{-\beta \mathsf{H}}$, the real time contour corresponding to \cref{eq:ModSkGen} with a complex $\bsig$ is shown. The region shaded yellow shows the image of time cylinder on the complex time plane. The marked points are identified to complete the \textit{thermal circle}. For this special case, both preparation of the state and its modular evolution is inbuilt into the choice of time contour via segments $C_{\textsf{in/out}}$ . Segments $C_{\sR/\sL}$ characterize evolution in real time and are required irrespective of the state. The origin on imaginary time circle is a matter of convention and is defined here with respect to the branch supporting anti time ordered evolution. Nevertheless, the path integral evolution is always towards negative imaginary time, starting from $t= i \beta \bsig$. }
	\label{fig:SKTimeCon}
	\end{center}
\end{figure}

Technically, the  difference between \cref{eq:TFDObs} and \cref{eq:SKObs} is essentially that in the former we have symmetrically arranged the thermal weight between the two sequences of temporally ordered operators corresponding to the dynamics in the entangled TFD state, while in the latter the full thermal weight is imposed at an initial time, giving correlations measured near equilibrium as the system evolves towards future. More generally, one can define a general class of correlators by distributing an arbitrary fraction of the density matrix, say $\bsig$ amount, across the time ordered sequence of operators (see \cref{eq:ModSkGen}). We will treat ingoing SK observables with the choice $\bsig=0$ as the basic objects and term other choices of observables as the modular transformed versions thereof. The parameter $\bsig$ will be termed the Euclidean modular time. In the equilibrium state all such choices of observables are equivalent since they can be easily analytically continued to each other \cite{kamenev_2011}. In the SK formalism, the value of $\bsig$ is defined by the choice of the time contour along which one performs the field theory path integral \cite{kamenev_2011}(see \cref{fig:SKTimeCon}).

However, away from equilibrium, it is not convenient to define correlators with fractionated insertions of the density operator. Even for states which are local excitations over equilibrium, the fractional density operator would in general have a complicated non-local description. Therefore the CGL prescription which directly derives ingoing SK observables from gravity is an important step towards developing a consistent paradigm of semi-classical near equilibrium dynamics of black holes, where one defines near equilibrium states as excitations over the grSK saddle.

Nevertheless, given that the value of $\bsig$ in the field theory can be freely decided in the equilibrium state, one would expect a similar choice to exist in the gravitational dynamics. In other words, we ask if there exists a simple CGL like prescription which defines a smooth bulk gravitational saddle corresponding to alternative choices of the boundary time contour. The main result of this work is an affirmative answer to this question. Our analysis demystifies a particular ingredient of the CGL prescription -- the grSK saddle was originally defined in \cite{Glorioso:2018mmw} as a specific analytical continuation of the Schwarzschild background parameterized in the ingoing Eddington-Finkelstein gauge along the radial direction. By moving away from this gauge restriction we show that the choice of different time slicings in the bulk translates to the choice of $\bsig$ in the boundary SK time contour. We identify a gravitational Wilson line like quantity we denote by $\totr{\csig}$, which could be thought of as the bulk dual of $\bsig$. More precisely, the value of $\bsig$ appear as the monodromy of $\totr{\csig}$ across the boundaries of the gravitational saddles we define (see \cref{eq:bsig}). The gauge invariance in the bulk is partially restored as many choices of gauge in the bulk result in identical expectation for this monodromy. In section \cref{sec:RindlerSKEW} we present a near horizon analysis giving a post facto justification for these results. As a dividend of this analysis, we are able to define a regular geometry which computes TFD correlators of the boundary, albeit with a complexified horizon. As evident from studies of the grSK geometry, this gives a convenient and efficient formalism to compute higher point TFD correlators of the boundary from gravity.

The CGL prescription is a prerequisite for the open effective theory (open EFT) paradigm for AdS black holes motivated in \cite{Jana:2020vyx, Loganayagam:2020eue}. The basic idea here is that the semi-classical effective dynamics around a black hole is dual to an open effective theory at the boundary -- the horizon is effectively a thermal reservoir which causes both dissipation through energy leakage and fluctuations via Hawking radiation. In the case of AdS black branes, this viewpoint cleanly extends the fluid gravity correspondence (FGC) \cite{Bhattacharyya:2007vjd,Hubeny:2011hd}, giving a bulk description of the fluctuating hydrodynamics of the boundary\cite{Ghosh:2020lel,He:2022jnc,He:2022deg}.\footnote{See \cite{Glorioso:2018mmw,deBoer:2018qqm} for alternate approaches to holographic fluctuating hydrodynamics.}  A less understood feature yet to be incorporated into this open EFT are the edge modes of the bulk gauge fields localized on the black hole horizon\cite{Koga:2001vq,Hotta:2000gx,Donnay:2015abr,Donnay:2016ejv,Chandrasekaran:2018aop, Chandrasekaran:2020wwn, Knysh:2024asf}. Edge modes are ubiquitous components of gauge theories on open manifolds which describe inequivalent gauge parameterizations of the boundary.  Therefore it is essential to accommodate them into the semi-classical open effective theory recorded by an exterior observer.\footnote{We emphasize that the edge modes we have in mind should not be confused with the long wavelength  quasinormal modes dual to hydrodynamic excitations.} Such edge modes have been argued to account for the Bekenstein entropy of the horizon \cite{Carlip:1993sa, Hotta:2002mq, Harlow:2016vwg, Mertens:2022ujr} (see also \cite{Hawking:2016msc}). This was one our inspirations to study the gauge dependence in the CGL prescription. Though we have not done a complete analysis of the edge modes in these geometries, our results show that a particular $0$ mode of the diffeomorphisms localized on the horizon of these geometries correspond to the choice of $\bsig$ (more precisely its real part) on the  boundary. In this sense, our analysis is a first step towards studying the open effective theory of the exterior which includes dissipation, fluctuations, edge modes and the interplay among them.

Finally, we will make some remarks related to recent works on the algebraic QFT  aspects of the bulk theory \cite{Witten:2021unn,Chandrasekaran:2022cip,Chandrasekaran:2022eqq,Penington:2023dql}. We will not attempt a rigorous comparison with these results. To state briefly, these studies adopt the TFD perspective on the bulk physics and argue that finite $G_{N}$ effects modify the algebra of bulk observables to a crossed product algebra of Type II nature \cite{Witten:2021unn}.  To the best of this author's understanding, these results imply that the bulk Hilbert space consists of states of the form
\begin{equation}\label{eq:CPstate}
	\begin{split}
	|\widetilde{\Psi}_{\beta,\mathsf{P}}\rangle &= \mathsf{P}(z)^{\frac{1}{2}} \otimes | \Psi; z \rangle_{_\text{TFD}} \ , \qquad  \qquad z \in \mathbb{R} \ , \\
	 	| \Psi; z \rangle_{_\text{TFD}} &= \mathcal{Z}_{\beta}^{-\frac{1}{2}}\sum_{n}  e^{-\left(\frac{1}{2}+i z\right)\beta E_{n}} | E_{n} \rangle_{_{\sR}} | E_{n} \rangle_{_{\sL}}  \ , \qquad \mathsf{P}(z)^{\frac{1}{2}} \in \mathbb{L}^{2}(\mathbb{R}) \ ,
	\end{split}
\end{equation}
and semi-classical excitations above them. The states $	|\widetilde{\Psi}_{\beta,\mathsf{P}}\rangle$ describe an entangled TFD pair of the CFT with a relative time shift $z$ between them,  where $z$ itself is drawn according to a probability distribution $\mathsf{P}(z)$. The corresponding boundary observables are given by
\begin{equation}\label{eq:CPObs}
	\text{Crossed Prod. Obs. :    }   \int \dd z \,  \mathsf{P}(z) X(z) \, \text{tr} \left\{ \left(\rho_{_{\beta}}\right)^{\frac{1}{2}+i z} \mathbb{T} \left( (\mathsf{O}_{1})_{_{\sR}} (\mathsf{O}_{2})_{_{\sR}} \ldots \right) \left(\rho_{_{\beta}}\right)^{\frac{1}{2}-iz} \widetilde{\mathbb{T}} \left( (\mathsf{O}_{1})_{_{\sL}} (\mathsf{O}_{2})_{_{\sL}} \ldots \right) \right\} ,
\end{equation}
where $X(z)$ is a bounded function of $z$. In the bulk, each state in this ensemble is represented by two sided geometries whose asymptotic boundaries are time shifted with respect to each other (see \cite{Maldacena:2013xja, Mandal:2014wfa, Papadodimas:2015xma} for earlier discussions). Again, we will not bother to specify a mathematically precise definition of the trace in \cref{eq:CPObs}. The expression \cref{eq:CPObs} is simply meant to characterize the possible correlations functions one can define at the boundary and their analytical structure.

The fact that the bulk semi-classical Hilbert space includes states of the form $	|\widetilde{\Psi}_{\beta,\mathsf{P}}\rangle$ naturally motivates the following question.  Had we taken the open EFT perspective to describe the bulk dynamics (or equivalently, worked with the grSK description of the bulk spacetime) what states capture the same physics as $|\widetilde{\Psi}_{\beta,\mathsf{P}}\rangle$? This is particularly confusing since in the ingoing SK description, the two boundaries of the spacetime do not represent two identical copies of the CFT, but rather the ket and bra degrees of freedom in the density operator of a single CFT. Therefore the analogue of $z$ should, naively, encode the relative translation between the ket and bra Hilbert spaces of a single system. In spite of this peculiar interpretation, we are required to consider such states seriously for the completeness and consistency of the open EFT paradigm.  Therefore, we suggest that the full Hilbert space of the open EFT of the black hole exterior should be enlarged to account for such ensembles of  modular time shifts (since $z = \Im \bsig$). We term the SK analogues of \cref{eq:CPObs} the \textit{SK ensemble observables}. In expression,
\begin{equation}\label{eq:CPObsSK}
	\text{SK ensemble obs. :}  \quad \int \dd z \,  \mathsf{P}(z) X(z)\, \text{tr} \left\{ \left(\rho_{_{\beta}}\right)^{i z} \mathbb{T} \left( (\mathsf{O}_{1})_{_{\sR}} (\mathsf{O}_{2})_{_{\sR}} \ldots \right) \left(\rho_{_{\beta}}\right)^{1-iz} \widetilde{\mathbb{T}} \left( (\mathsf{O}_{1})_{_{\sL}} (\mathsf{O}_{2})_{_{\sL}} \ldots \right) \right\} .
\end{equation}
In \cref{sec:ModAvgStates} we argue that once we allow for analytically continued gravitational saddles \`{a} la CGL, averaged boundary observables as in \cref{eq:CPObsSK} are generated naturally from the bulk. 

We conclude this section with an outline of the paper. In \cref{sec:ModFlowQFT} we begin by reviewing basic properties of Euclidean modular flowed SK correlators in a field theory. In \cref{sec:BBSKEW} we demonstrate how CGL like analytical continuations of a black hole generates Euclidean modular flowed SK correlators from gravity. As an illustrative example, we work with the black brane solution and derive $2$-pt correlations functions of a probe Klein-Gordon field. In \cref{sec:RindlerSKEW}
we give a near horizon analysis justifying these results. In particular, we also clarify the connection between these geometries and the usual eternal geometries, and how our construction follows from the JLMS relation. The emergence of SK ensemble observables is discussed in \cref{sec:ModAvgStates}. Finally, in \cref{sec:KScond} we show that the geometries we define satisfy a real time version of the Kontsevich-Segal consistency conditions.

\section{Modular flowed real time correlators}\label{sec:ModFlowQFT}

Consider a quantum system and the density operator $\rho$ describing its state. The generating functional for modular flowed  real time or Schwinger-Keldysh (SK) correlators in this state is defined as
\begin{equation}\label{eq:ModSkGen}
	\mathcal{Z}[J_{\sR},J_{\sL};\bsig] := \tr \left\{ \left(\rho \right)^{\bsig} \mathsf{U}[J_{\sR}]  \, \left(\rho \right)^{1-\bsig} \, \mathsf{U}^{\dagger}[J_{\sL}] \right\} \ ,  \quad \rho:= e^{- 2\pi \mathsf{K}}
\end{equation}
where $\mathsf{K}$ is the modular Hamiltonian corresponding to $\rho$ and $\bsig$ is the Euclidean modular time (which can be complex). In general, the modular Hamiltonian $\mathsf{K}$ could be a non-local operator. Since all physical states are defined by a positive and Hermitian $\rho$, it is sensible to imagine it as the exponential of some $\mathsf{K}$. $\mathsf{U}[J]$ denotes the time ordered unitary evolution operator
\begin{equation}
	\mathsf{U}[J] : = \mathbb{T} \,  \text{exp} \left[- i \int \dd t  \left(\mathsf{H} + J \mathsf{O} \right)   \right] \ .
\end{equation}
$\mathsf{H}$ denotes the Hamiltonian of the system. In the particular case of a thermal state $\rho = \rho_{_\beta}$, we have $\mathsf{K}= \frac{\beta}{2 \pi} \mathsf{H}$. In this convention, the modular time  $\bsig$ for a thermal state is defined with an implicit scale set by the inverse temperature $\beta$. 

The source $J$ and operator $\mathsf{O}$ are in general local, and their spatial dependence has been suppressed in the above schematic expression. In \cref{eq:ModSkGen} we have denoted sources defining the evolution towards future and past as $J_{\sR}$ and $J_{\sL}$ respectively. When we discuss correlators, the corresponding insertion of $\mathsf{O}$  will be denoted as $\mathsf{O}_{\sR/\sL}$ (thought of as \textit{right}/\textit{left} versions) respectively. $\mathsf{O}$  is taken to be Bosonic and Hermitian  (generalization to Fermionic case is straightforward with minor modifications related to Grassmann multiplication).

Let us now be more explicit about the correlators under consideration.  Correlation functions can be generated from \cref{eq:ModSkGen} by acting with $\mathsf{O}_{\sR/\sL} \sim \frac{\pm}{i} \frac{\delta }{\delta J_{\sR/\sL}}$ and are of the form
\begin{equation}\label{eq:ModNpt}
	\begin{split}
		\langle \mathsf{O}_{a_{1}}(t_{1})  \mathsf{O}_{a_{2}}(t_{2}) \cdots \mathsf{O}_{a_{n}}(t_{n})\rangle (\bsig) &:=  \left\langle  \left(\rho\right)^{-\bsig} \left( \widetilde{\mathbb{T}} \! \! \!  \prod_{i=1, \; a_{i}=\sL}^{n}  \! \! \!  \mathsf{O}(t_{i}) \right) \left(\rho\right)^{\bsig}\left( \mathbb{T}  \! \! \!  \prod_{j=1, \;  a_{j}=\sR}^{n} \! \! \! \mathsf{O}(t_{j}) \right) \right\rangle_{\rho} \ , \\
		& = \left\langle   \left(\rho\right)^{\bsig-1}\left( \mathbb{T}  \! \! \!  \prod_{j=1, \;  a_{j}=\sR}^{n} \! \! \! \mathsf{O}(t_{j}) \right)  \left(\rho\right)^{1-\bsig} \left( \widetilde{\mathbb{T}} \! \! \!  \prod_{i=1, \; a_{i}=\sL}^{n}  \! \! \!  \mathsf{O}(t_{i}) \right) \right\rangle_{\rho} \ ,
	\end{split}
\end{equation}
where $\langle \dots \rangle_{\rho} := \tr \left(\rho \ldots \right)$ and $\mathbb{T}$ ($\widetilde{\mathbb{T}}$) is the (anti) time ordering symbol. 
SK correlators are built of two on-overlapping strings of time ordered $\sR$ operators and anti time ordered $\sL$ operators. The modular flow results in a relative shift of left  operators along the modular time (which for the thermal state is same as Euclidean time). Alternatively, cyclicity of the trace operation allows us to view this as the shift of ket operators backward in modular time. In general, the modular Hamiltonian $\mathsf{K}$ is positive but unbounded from above, which could render the  $\rho_{_{\beta}}^{\pm \bsig}$ factors in \cref{eq:ModNpt} singular. Therefore the real part of $\bsig$ should be restricted as
\begin{equation}\label{eq:SigBoundQFT}
	\Re \bsig \in (0,1) \ .
\end{equation}
In the simplest case of $2$-pt functions, the correlators defined here are given by
\begin{equation}\label{eq:Mod2pt}
	\begin{split}
		\langle  \mathsf{O}_{\sR}(t_{1}) \mathsf{O}_{\sR}(t_{2}) \rangle(\bsig) & = \langle \mathbb{T} \, \mathsf{O}(t_{1}) \mathsf{O}(t_{2}) \rangle_{\rho} \ , \qquad \langle \mathsf{O}_{\sR}(t_{1}) \mathsf{O}_{\sL}(t_{2})\rangle (\bsig)= \langle e^{2\pi \bsig \mathsf{K}}\mathsf{O}(t_{2}) e^{-2\pi \bsig \mathsf{K}}\mathsf{O}(t_{1}) \rangle_{\rho} \ , \\
		\langle \mathsf{O}_{\sL}(t_{1}) \mathsf{O}_{\sR}(t_{2})\rangle (\bsig)&= \langle e^{2\pi \bsig \mathsf{K}}\mathsf{O}(t_{1}) e^{-2\pi \bsig \mathsf{K}}\mathsf{O}(t_{2}) \rangle_{\rho} \ , \qquad  \langle  \mathsf{O}_{\sL}(t_{1}) \mathsf{O}_{\sL}(t_{2}) \rangle (\bsig)  = \langle \widetilde{\mathbb{T}} \, \mathsf{O}_{\sL}(t_{1}) \mathsf{O}_{\sL}(t_{2}) \rangle_{\rho}  \ .
	\end{split}
\end{equation}
The modular flow has non-trivial effect only on the correlations between the right type and left type operators. It does not modify purely right or left type correlations. The two limit cases $\bsig=0$ and $\bsig =1$ have physically interesting interpretations. The case $\bsig=0$ gives the usual real time correlations describing evolution of  the initial state $\rho$ towards the future. Such correlators are more familiar and extensively studied in the context of response theory, particularly near thermal equilibrium \cite{kamenev_2011}. The case $\bsig=1$ defines real time correlations in a system which is approaching the final state $\rho$.  We will refer to the correlations with $\bsig=0$ and $\bsig=1$ as \textit{ingoing frame} and \textit{outgoing frame} correlations respectively. For two sets of operators  $ \mathsf{A}_{i}$ and $ \mathsf{B}_{j}$, we have

\begin{equation}
	\begin{split}
		\Big\langle \prod (\mathsf{A}_{i})_{\sR} \, \prod (\mathsf{B}_{j})_{\sL} \Big\rangle_{\text{in}} &= \Big\langle \widetilde{\mathbb{T}}\left(\prod \mathsf{B}_{j}\right)\,  \mathbb{T}\left(\prod \mathsf{A}_{i} \right) \Big\rangle_{\rho} \ ,  \\
		\Big\langle \prod (\mathsf{A}_{i})_{\sR} \, \prod (\mathsf{B}_{j})_{\sL} \Big\rangle_{\text{out}} &= \Big\langle   \mathbb{T}\left(\prod \mathsf{A}_{i} \right) \,  \widetilde{\mathbb{T}}\left(\prod \mathsf{B}_{j}\right) \Big\rangle_{\rho} \ ,
	\end{split}
\end{equation}

More generally, the relative ordering of the ket operators and bra operators in a correlator is exchanged under the transformation $\bsig \to 1- \bsig$ as shown in the second line of \cref{eq:ModNpt}. The invariant point under this action given by $\bsig = \frac{1}{2}$ is termed the Thermo Field Double (TFD) configuration. TFD correlations are given by
\begin{equation}\label{eq:TFDComm}
	\Big\langle \prod (\mathsf{A}_{i})_{\sR} \, \prod (\mathsf{B}_{j})_{\sL} \Big\rangle_{_\text{TFD}} = \Big\langle \rho^{-\frac{1}{2}}\,\widetilde{\mathbb{T}}\left(\prod \mathsf{B}_{j}\right)\,\rho^{\frac{1}{2}}\,  \mathbb{T}\left(\prod \mathsf{A}_{i} \right) \Big\rangle_{\rho} = \Big\langle   \rho^{-\frac{1}{2}}\,\mathbb{T}\left(\prod \mathsf{A}_{i} \right) \, \rho^{\frac{1}{2}}\,  \widetilde{\mathbb{T}}\left(\prod \mathsf{B}_{j}\right) \Big\rangle_{\rho}\ .
\end{equation}

\subsection{Thermal states and KMS relations}

We now specialize to the case $\rho = \rho_{_\beta}$, where $\rho^{\bsig} \propto e^{-\beta \bsig \mathsf{H}}$. The modular evolution via $\mathsf{H}$ reduces to temporal evolution in Euclidean time direction. As a consequence the thermal correlators satisfy Kubo-Martin-Schwinger (KMS) relations \cite{doi:10.1143/JPSJ.12.570,PhysRev.115.1342}. These relations can be expressed effectively in terms of the frequency space operators defined via
\begin{equation}
	\mathsf{O}(t) = \int \frac{\dd \w}{2 \pi} \,  \mathsf{O}(\w) e^{-i \w t} \ 
\end{equation}
as,
\begin{equation}
	\big\langle \mathsf{O}(\w_{1}) \mathsf{O}(\w_{2}) \ldots \mathsf{O}(\w_{n})  \big\rangle_{\rho_{_{\beta}}} = e^{\beta \w_{1}}\big \langle \mathsf{O}(\w_{2}) \ldots \mathsf{O}(\w_{n}) \mathsf{O}(\w_{1}) \big \rangle_{\rho_{_{\beta}}}  \ .
\end{equation}
Relations such as above relate correlators with cyclic re-orderings of the operator insertions (here we have assumed Bosonic statistics and the Fermionic counterpart will have additional statistical phase factors). At the level of two point functions the above KMS structure reduces all correlations to a single correlator. Consider the retarded correlator defined as
\begin{equation}
	\begin{split}
	\langle \mathsf{O}(t_{1}) \mathsf{O}(t_{2}) \rangle^{_\text{Ret}} &:= \theta(t_{1}-t_{2})\langle \left[\mathsf{O}(t_{1}), \mathsf{O}(t_{2}) \right] \rangle_{\rho_{_\beta}}  = \int \frac{\dd \w}{2 \pi} \,  G^{_\text{Ret}}(\w) e^{-i \w (t_{1} -t_{2})} \ .
	\end{split}
\end{equation}
The KMS relations for $2$-pt correlations then give
\begin{equation}
2  \nb  \Re  G^{_\text{Ret}}(\w) =  \langle  \mathsf{O}(-\w )\mathsf{O}(\w) \rangle_{\rho_{_\beta}} \ ,
\end{equation}
with the Bose-Einstein weight $\nb$ defined as\footnote{We suppress the frequency dependence in $\nb \equiv \nb(\w)$ and repackage all instances of $\nb(-\w)$ using the relation $\nb(\w) + \nb(-\w) + 1=0$. \label{fn:nBRel}}
\begin{equation}
\nb := \frac{1}{e^{\beta \w} -1} \ .
\end{equation}
It is straightforward to verify that the correlators in \cref{eq:Mod2pt} for the thermal state  then satisfy\footnote{Alternatively, 
\begin{equation}
\langle \mathsf{O}_{\sL}(\w) \mathsf{O}_{\sL}(-\w)\rangle = \left(G^{_\text{Ret}}(\w)\right)^{*} + 2  \nb \Re G^{_\text{Ret}}(\w) \ .
\end{equation}
}
\begin{equation} \label{eq:ModKMSRel}
\begin{split}
\langle \mathsf{O}_{\sR}(\w) \mathsf{O}_{\sR}(-\w)\rangle &= G^{_\text{Ret}}(\w) + 2  \nb \Re G^{_\text{Ret}}(\w) \ , \\
\langle \mathsf{O}_{\sR}(\w) \mathsf{O}_{\sL}(-\w)\rangle &=   2 \nb \, e^{\bsig \beta  \w }  \Re G^{_\text{Ret}}(\w) \ , \\ 
\langle \mathsf{O}_{\sL}(\w) \mathsf{O}_{\sR}(-\w)\rangle &=   2 (\nb+1) \, e^{-\bsig \beta  \w }  \Re G^{_\text{Ret}}(\w) \ , \\ 
\langle \mathsf{O}_{\sL}(\w) \mathsf{O}_{\sL}(-\w)\rangle &= -G^{_\text{Ret}}(\w) + 2  (\nb+1) \Re G^{_\text{Ret}}(\w) \ .
\end{split}
\end{equation}

The generating functional in \cref{eq:ModSkGen} can be defined as a SK path integral on a time contour which evolves fields forward and backward selectively \cite{kamenev_2011}. In particular, for the thermal state, both the preparation of the density matrix and its modular flow can be inbuilt into this path integral as part of the time contour (see \cref{fig:SKTimeCon}).

In the next section we discuss the gravitational dual of modular flowed SK correlations and re-derive the relations \cref{eq:ModKMSRel} in a bulk semi-classical description.

\section{Schwinger-Keldysh eternal wormholes}\label{sec:BBSKEW}

Thermal states in holographic CFTs are well understood as dual to blackholes in AdS spacetimes (more precisely, we study systems above the Hawking-Page phase transition). It is a natural question to ask how should one understand the modular flowed SK correlators defined in the last section from a purely bulk perspective. In this section we show that such a description is possible via a straight forward generalization of the real time prescription in \cite{Glorioso:2018mmw}. As an illustrative example, we will focus on the Schwarzchild black brane in asymptotically AdS$_{d+1}$ spacetime.

We parameterize the black brane as
\begin{equation}\label{eq:BGmetric}
 \dd s^{2}= \gorb_{\alpha \beta}\dd y^{\alpha} \dd y^{\beta} + r^{2}\dd \bx^{2} \ ,
\end{equation}
where $y=\{w,r\}$ denotes coordinates for the $2$ dimensional \emph{orbit space} of the black brane, and $\bx$ run along the brane's world volume $\mathbb{R}^{d-1}$ plane directions. The orbit space metric $\gorb$ is parameterized in a novel manner:
\begin{equation} \label{eq:sigmet}
\begin{split}
	\gorb	
%	=&  -r^{2} f(br) \, \dd w  \left[\dd w- 2\sig(r)  \fr{\dd r}{r^{2}f(br)} \right] + (1-\sig^{2})\fr{\dd r^{2}}{r^{2}f(br)}   \\
	=& - r^{2} f(br) \left( \dd w + \frac{1- \sig(r) }{r^{2}f(br) } \dd r\right)   \left(\dd w - \frac{1 + \sig(r)}{r^{2}f(br) } \dd r\right)  \ ,  \qquad f(y) = 1 - y^{-d} \ .
	\end{split}
\end{equation}
We will henceforth often suppress the arguments of functions $f \text{ and } \sig$. The inverse Schwarzschild radius  $b=r_{h}^{-1}$ is related to the inverse temperature  as $\beta = \frac{4 \pi b}{d}$.  $\sig(r)$ parameterizes choice of spatial slicings of the spacetime  invariant under boundary translations. 
Its range is restricted to satisfy
\begin{equation} \label{eq:Sigbound}
\sig(r) \in (-1,1) \ ,
\end{equation}
which ensures that fixed $w$ hypersurfaces are spatial slices.\footnote{In \cref{sec:KScond} we show that, near the horizon, this restriction is essential to satisfy the Kontsevich-Segal consistency conditions.} The metric in \cref{eq:sigmet} is equivalent to the usual Schwarzschild metric with time $w_{s}$ via the diffeomorphism 
\begin{equation}\label{eq:DiffToSch}
 \dd w  \to \dd w_{s}+ \sig \frac{\dd r}{ r^{2}f  }  \ , \qquad  \dd r \to \dd r \ . 
\end{equation} 
Notice that setting $\sig= \left\{ 0 ,  \pm 1 \right\}$ give the Schwarzschild and ingoing/outgoing  Eddington-Finkelstein coordinates respectively. We may also think of \cref{eq:sigmet} as a family of coordinate systems for the black brane  generated by the flow
\begin{equation}\label{eq:SigFlow}
 \gorb_{\alpha \beta}(\sig)  =  \flow(\sig)_{\alpha}^{\ \gamma} \flow(\sig)_{\beta}^{\ \delta} \, \mathtt{g}_{\gamma \delta}^{(0)} \ , \qquad \flow_{\alpha}^{\ \beta}(\sig) = \begin{bmatrix}
                 1 & 0 \\
                 -\frac{\sig}{r^{2}f}& 1 
                  \end{bmatrix} \ .
\end{equation}

\begin{figure}[h!] 
	\begin{center}
		\begin{tikzpicture}[scale=0.6]
			\draw[thin, color=teal!50,  ->] (-8,-2) -- (-8,2) node [above] {$\scriptstyle{\Im r}$};
			\draw[thick,color=teal!50,fill=teal!50] (-8,0) circle (0.35ex);
		\draw[thin, color=teal!50,  -] (-10,0) -- (-5,0); 	
		\draw[thin, color=teal!50,  ->] (5,0) -- (7.5,0) node [right] {$\scriptstyle{\Re r}$};  
		\draw[thin, snake it, color=teal!50] (-8,0) -- (-10,-0.5);
		
			\draw[thin,color=black] (-5,0) circle (0.05ex) node[below] {$\scriptstyle{r_h}$};
			\draw[thick,color=teal!70,fill=teal!70] (-5,0) circle (0.35ex);
			\draw[thin,snake it, color=teal!70] (-5,0)  -- (5,0) ;
		
			\draw[thick,color=olive!50!teal,fill=olive!50!teal] (5,1) circle (0.35ex);
			\draw[thick,color=olive!50!teal,fill=olive!50!teal] (5,-1) circle (0.35ex);
			\draw[thick,color=olive!50!teal, ->-] (5,1)  node [right] {$\scriptstyle{r_c+i 0 }$} -- (0,1) node [above] {$\scriptstyle{\alpha < 0, \ \Re \ctor =0}$} -- (-4,1);
			\draw[thick,color=olive!50!teal,->-] (-4,-1) -- (0,-1) node [below] {$\scriptstyle{\alpha>0 , \ \Re \ctor =1}$} -- (5,-1) node [right] {$\scriptstyle{r_c-i 0}$};
			\draw[thick,color=olive!50!teal,->-] (-4,1) arc (45:315:1.414);
		\end{tikzpicture}
		\caption{ The complex $r$ plane at a fixed $w$: the SKEW contour is a codimension-1 surface in this plane with a cut off radius $r_c$. As indicated, the direction of contour is counter-clockwise around the branch point at the horizon where the branch cut of $\log f$ is anchored. Any other relevant branch cut (such as that of $\log r$) is placed away from the SKEW contour.}
		\label{fig:mockt}
	\end{center}
\end{figure}
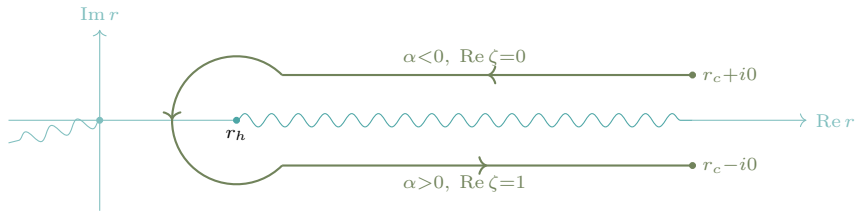

Let us now proceed to define the Schwinger-Keldysh (SK) avatar of the geometry in \cref{eq:BGmetric}. We complexify the metric in \cref{eq:BGmetric} following \cite{Glorioso:2018mmw}, by analytically continuing the radial coordinate to a $1$-dimensional contour on the complex $r$ plane. The contour can be parameterized using the  real variable $\alpha \in (-1,1)$, which is related to the radial coordinate via
\begin{equation}\label{eq:AlphaDef}
	f(\alpha) :=  \left(\alpha - i 0 \right)^{2} \ , \quad  r(\alpha) := r_{h} \left( 1 - f(\alpha)  \right)^{- \idim}  \ ,
\end{equation}
where $0$ denotes an infinitesimal positive real number and $\idim := d^{-1}$.  Clearly, we should select the branch of $r(\alpha)$ close to the positive real line.\footnote{More generally, the complexification considered here analytically continues the lapse field to negative values. For asymptotically AdS spacetimes as in the present case, a convenient choice is $ r^{-2} \gorb_{w w} = - (\alpha - i 0)^{2}$.} Using $\frac{\dd r}{\dd \alpha} =: r'$, the SK version of the orbit space metric is given by
\begin{equation} \label{eq:AlphaMet}
	\begin{split}
		\gorb_{_\text{SK}}
		=& - r^{2} f \left( \dd w + \frac{1- \sig }{r^{2}f} r' \dd \alpha \right)   \left(\dd w - \frac{1 + \sig}{r^{2}f } r'  \dd \alpha\right)  \ , \quad r' = 2 \idim \, b^{d} \, r^{d+1} (\alpha- i 0) \ ,
	\end{split}
\end{equation}
where all quantities now dependent on $\alpha$. We refer to complexified black brane geometries with the orbit space  metric $\gorb_{_\text{SK}}$ as  \textit{Schwinger-Keldysh eternal wormhole} (SKEW) geometries. The resulting radius contour on the complex plane is shown in \cref{fig:mockt}. The segments of the contour with positive and negative imaginary radius correspond to negative and positive $\alpha$ respectively.\footnote{\label{fn:alpha0} We can adjust the imaginary shift in $\alpha$ and deform the resultant $r$ contour, by taking $\alpha - i 0 \to \alpha - i 0\, h(\alpha)$, where $h$ is an even function of $\alpha$. For example, defining $h(\alpha) = (1-\alpha^{2})^{2}$ gives a radial contour nearer to the real $r$ line for large $\Re r$. Similarly, an asymptotic cut off radius can be introduced by restricting $\alpha$ to $(-\alpha_{c}, \alpha_{c})$.} 

Though the above parameterization using $\alpha$  is important (especially for the analysis in \cref{sec:KScond}), we now give a more practical description of the same geometry.  As seen from studies of the grSK saddle, it is convenient to parameterize this geometry using a mock tortoise radius coordinate which has a logarithmic branch point at the horizon. In this regard, we define the \emph{tortoise transform} of a function $\mathcal{F}(r)$ as
\begin{equation}\label{eq:tortrans}
	\totr{\mathcal{F}}(r) :=  \frac{2}{i\beta} \int\displaylimits_{\infty+i 0}^{r}\frac{\mathcal{F}(y) \dd y}{ y^{2} f(by)}  =  \frac{d}{2 \pi i }\int \displaylimits_{\infty+i 0}^{b r} \mathcal{F}\left(t/b\right) \frac{t^{d-2}}{t^d -1} \dd t  \ .
\end{equation}
A mock tortoise coordinate can be defined on the geometry via 
\begin{equation}
	\ctor(r) := \totr{1} \ .
\end{equation}
The tortoise transformed functions defined as above typically have logarithmic branch cuts at the horizon $br =1$, which essentially descend from that of $\log f$ (see \cref{fig:mockt}). We depict the branch cut to lie along the ray $[r_{h},\infty)$.  With this convention, when continued around the horizon to below the cut, such functions typically undergo monodromy shifts.   As an example, for a function $\mathcal{F}$ otherwise analytic at the horizon, its tortoise transform $\totr{\mathcal{F}}$ undergoes a shift upon continuing around the branch point as
\begin{equation}\label{eq:TorMono}
	\totr{\mathcal{F}}(r - i 0) = \totr{\mathcal{F}}(r + i 0) + \mathcal{F}(r_{h}) \ .
\end{equation}
The normalization in \cref{eq:tortrans} has been fixed to set the monodromy shift in $\totr{\mathcal{F}}$ to the value of $\mathcal{F}$ at the horizon. In particular, we have
\begin{equation}
\ctor(r - i 0) - \ctor(r + i 0) = 1 \ . 
\end{equation}

The SKEW geometry can be thought of as a complexified spacetime defined by the analytical continuation of \cref{eq:sigmet}  on to a radial contour parameterized by $\ctor$ -- at every fixed $w$ time slice, the radial contour approaches the horizon from the asymptotic infinity $r = \infty + i 0$, turns around the branch cut at $r_{h}$ and continues back to the asymptotic infinity at $\infty - i 0 $ (see \cref{fig:mockt}).  The top and bottom exterior segments of the contour can be distinguished by noting that they have $\Re \ctor = 0$ and $1$ respectively. The circular segment of the radial contour which winds around the branch point is addressed as the \textit{horizon cap}, which corresponds to the region $|\alpha| < \alpha_{h} $ for some small positive $\alpha_{h}$ (see \cref{fn:alpha0}).

\begin{figure}[h!]
	\begin{center}
		\includegraphics[scale=0.20]{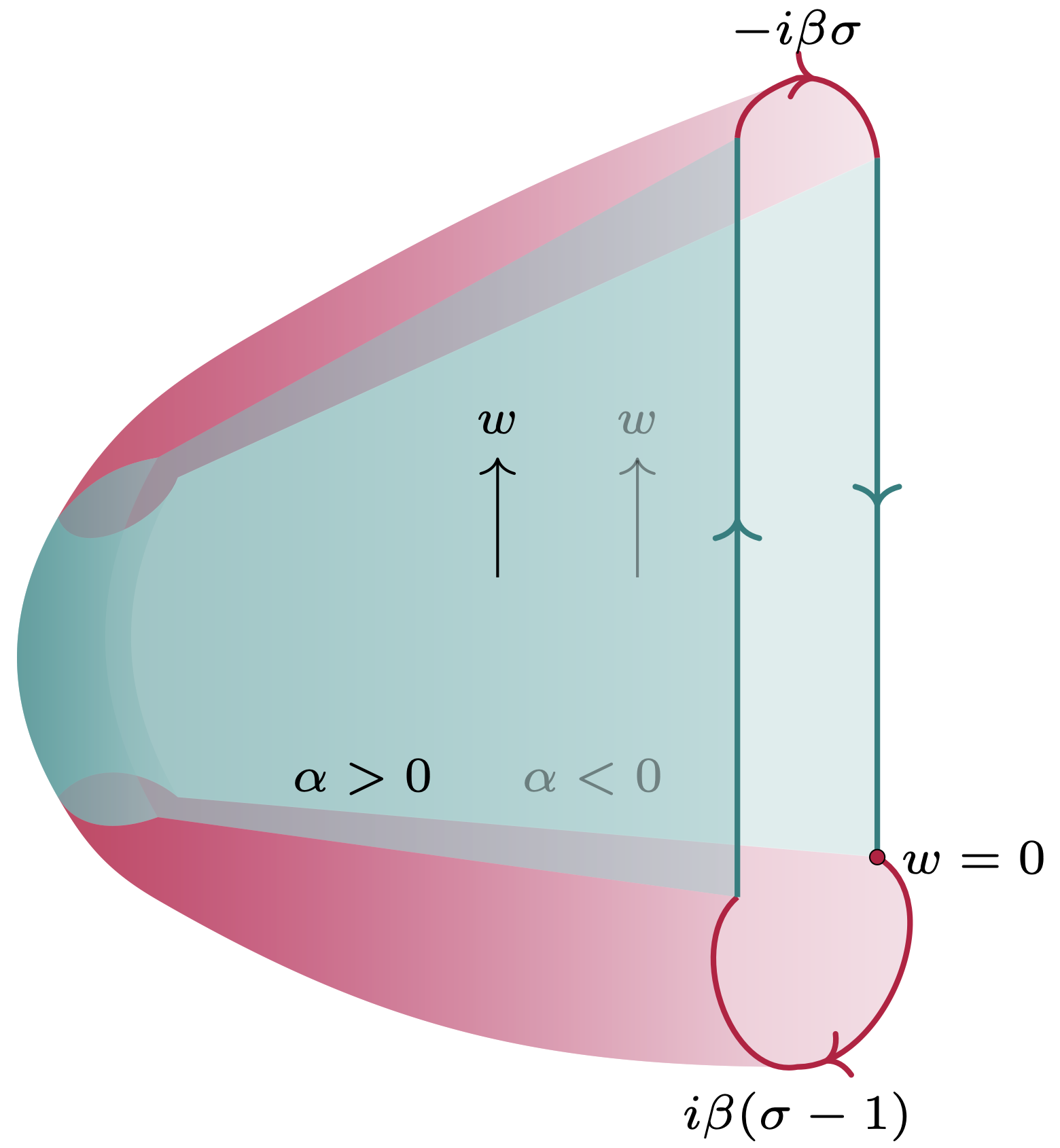} 
	\end{center}\caption{The schematic structure of a completed SKEW geometry is shown. Lorentzian region  covered by the radial contour is shown in teal, with additional Euclidean caps (in purple) attached to the past and future most time slices. The Euclidean caps are complementary sections of the cigar geometry. The asymptotic boundary matches that of a modular flowed SK contour, with the contour time flow marked in respective colors.}
\label{fig:GGSKbulk}
\end{figure}

As $\sig$ appears in \cref{eq:sigmet} as only a coordinate redefinition, usually one need not distinguish between various metrics on the flow trajectory \cref{eq:SigFlow}. However, once complexified and analytically continued to a SKEW geometry, metrics with distinct choices of $\sig$ correspond to inequivalent thermal contours of the dual CFT. In \cref{sec:Probes}, this claim is evidenced by probing the black brane with a scalar field.  Before proceeding further, let us define
\begin{equation}
	\csig(r) := \frac{1}{2}(1-\sig(r))  \ .
\end{equation}
The shifted version of $\sig$ given by $\csig$ will prove more convenient to characterize field perturbations in the spacetime. In terms of its tortoise transform $\totr{\csig}$, the coordinate $w$ can be mapped to the ingoing time $w_{in}$ via the diffeomorphism
\begin{equation}\label{eq:DiffToIn}
w \to  w_{in} - i \beta \totr{\csig} \ , \qquad r \to r \ .
\end{equation}
We will show that holographic correlators generated from the SKEW spacetime correspond to modular flowed correlations with
\begin{equation}\label{eq:bsig}
	 \bsig := \lim_{r \to \infty-i 0} \totr{\csig}(r) = \bbsig +i \isig \ ,
\end{equation}
were we have allowed the modular flow to be complex in general, with real part  $\bbsig$ and imaginary part $\isig$.\footnote{In terms of the bulk ADM decomposition $\bsig$ may be related to the $0$-mode of the radial shift vector on the full geometry. A more accurate description is in terms of a \textit{radial ADM decomposition} where $\sigma$ quantifies the total shift of the radial slice along the time direction across the asymptotic boundaries of the SKEW geometry. \label{fn:RadialADM}} This is the main proposal of this work.  We will not provide a top-down proof for this claim, but rather provide evidence for it akin to studies of the grSK saddle, by constructing simple boundary correlators in \cref{sec:Probes} and verifying that they have the expected Kubo-Martin-Schwinger (KMS) structure. A schematic diagram of the geometry is shown in \cref{fig:GGSKbulk}.

\subsection{Constructing bulk time slices}\label{sec:SigProf}

Given that the precise value of $\bsig$ in the field theory can be freely decided, we expect this choice in the bulk description as well.   In \cref{eq:bsig}, $\bsig$ is defined as the monodromy shift gained by $\totr{\csig}$ upon traversing the whole radial contour.  Therefore,  each boundary thermal contour parameterized by $\bsig$ is represented in the bulk by a family of contours parameterized by $\csig$ and related via residual gauge symmetries which leave $\bsig$ invariant. In particular, how the monodromy is  generated from the bulk is irrelevant. However, our means to generate the monodromy is restricted by \cref{eq:Sigbound}, which ensures that the spacetime is foliated spatially with respect to $w$. It forces us to avoid any $\csig$ which is singular (in the sense of a pole or branch cut) on the exterior spacetime.

If we choose $\csig$ to be both analytic and real on the exterior, we could as well read out the  real part $\bbsig$ as a localized contribution from the horizon as in \cref{eq:TorMono}. In expression,
\begin{equation}
	\sig \text{   analytic on the contour:    }   \qquad   \bbsig = \csig(r_{h}) \ ,  \qquad \isig = 0 \ .
\end{equation}
 The spatial slicing condition in \cref{eq:Sigbound} translates to 
\begin{equation} \label{eq:ReSigBound}
	\csig(r) \in (0,1) \ , \qquad \bbsig \in (0,1) \ ,
\end{equation}
naturally reproducing \cref{eq:SigBoundQFT}.  For such a choice, the real part of modular time given by $\isig$ vanishes. It would seem that to generate  correlators with a non-trivial $\isig$, we need to perform a computation with the above choice and then analytically continue $\bbsig$ to the complex plane. Remarkably, there exists a special class of non-analytic functions which are allowed as possible $\sig$ (equivalently $\csig$). Consider the class of functions
\begin{equation}
	\sig = \sig_{\av} +\sig_{\dif}  \frac{ \sqrt{f}}{2} \ ,
\end{equation}
where $\sig_{\av}$ and $\sig_{\dif}$ are analytic  functions of $r$ on the full SKEW contour (and a finite sliver of the complex $r$ plane around the contour). $\sig$ defined as above is never singular except for a quadratic branch cut at the horizon inherited from $\sqrt{f}$.  In addition, in the exterior region, it is always real, a property which fails for  generalizations involving other fractional powers of $f$. The above form with $\sqrt{f}$ as the dressing factor is a representative for any $\sig$ with quadratic branch cut on SKEW contour. We work with the branch where $\sqrt{f}$ is negative above the branch cut, and satisfies
\begin{equation}
	\sqrt{f(br-i 0)} -\sqrt{f(br+i 0)}  =  2 \left| \sqrt{f(br)} \right| \ .
\end{equation}
This matches with \cref{eq:AlphaDef} as $\sqrt{f} = \alpha -i 0$.  The corresponding version of $\csig$ is defined as 
\begin{equation}
	\csig  = \csig_{\av} + \csig_{\dif}  \frac{ \sqrt{f}}{2}  \ , \qquad   \csig_{\av} :=  \frac{1-\sig_{\av}}{2}  \ , \qquad \csig_{\dif} := -\frac{\sig_{\dif}}{2} \ .
\end{equation}
Note that analytic profiles of $\sig/\csig$ with respect to $r$ translate to profiles which are even functions of $\alpha$ and profiles with square root branch cut correspond to odd functions of $\alpha$. Calculating $\bsig = \bbsig + i\isig$ gives
\begin{equation} \label{eq:RsigIsig0}
	\bbsig = \csig_{\av}(r_{h}) \ , \qquad \isig = -\frac{i}{2} \lim_{r \to \infty -i 0} \totr{\left(\csig_{\dif} f^{\frac{1}{2}} \right)}   =  - \frac{d}{2 \pi} \int \displaylimits_{1}^{\infty}    \csig_{\dif}(t/b) \frac{t^{\frac{d}{2}-2}}{\sqrt{t^{d}-1}} \dd t \ .
\end{equation}
To compute the above, we have used that since $\csig_{\av}$ is analytic, it contributes only to the real part of $\bsig$ as a residue from the horizon branch point. Further, there is no such residue contribution from $\csig_{\dif}$ as it multiplies a $\sqrt{f}$ factor which vanishes at the horizon. In contrast, $\csig_{\dif}$ contributes to the imaginary part $\isig$ as a non-trivial discontinuity integral, again due to the branch cut in its accompanying $\sqrt{f}$ factor. In \cref{eq:RsigIsig0} we have expressed this discontinuity integral as a single integral in the exterior.

Recall that $\sig_{\av/\dif}$ are not arbitrary, and from \cref{eq:Sigbound} satisfy
\begin{equation}
	|\sig_{\av}| + |\sig_{\dif} |  |\sqrt{f}|  < 1 \ ,
\end{equation}
which at first sight seems to further restrict the range of $\bbsig$ when we turn on a non-trivial $\isig$. To remedy this, we can work with a $\sig$ which interpolates between different behaviours and maximize both $\bbsig$ and $\isig$ independently. As an example, consider 
\begin{equation}\label{eq:SigInter0}
	\sig = (1-2\tau) \mathbb{1}_{h}+ s \sqrt{f}  \left( 1- \mathbb{1}_{h} \right) \ ,
\end{equation}
where  $\mathbb{1}_{h}$ is an indicator function which evaluates to $1$ on the horizon cap and $0$ elsewhere on the contour. $\tau $  and $s$ are taken to be constants for now. The function $\mathbb{1}_{h}$ serves to interpolate $\sig$ across the stretched horizon region such that 
\begin{equation} \label{eq:SigInter}
	\sig  \approx  \Bigg\{ \begin{array}{c}
		1-2\tau    \quad \text{on the horizon cap, }  \\
		s \sqrt{f}  \quad \text{far from horizon, } 
	\end{array}  \quad   \qquad \csig  \approx  \Bigg\{ \begin{array}{c}
		\tau    \quad \text{on the horizon cap, }  \\
		- \frac{s}{2} \sqrt{f}  \quad \text{far from horizon.} 
	\end{array} 
\end{equation}
It is possible to be more explicit and construct  an  analytical function $\mathbb{1}_{h}$ on the contour with the above features as following. Consider the horizon cap on the $r$ contour as a fixed curve encircling the branch point $br=1$. Then on the horizon cap we have 
\begin{equation}
 \Re f < \varepsilon  \ , \qquad \varepsilon := \max_{r \in \text{hor. cap}} |f|  \ .  
\end{equation}
Alternatively, $\varepsilon$ can be fixed as the value of $|f|$ at the stretched horizon.  A suitable choice for $\mathbb{1}_{h}$  is given by
\begin{equation}
\mathbb{1}_{h}  :=  \frac{1}{e^{\Lambda (f-\varepsilon)} +1 } \ , \qquad \Lambda  \gg \varepsilon^{-1} \ .
\end{equation}
For large $\Lambda$, the combination $e^{\Lambda (f-\varepsilon)}$ blows up in the exterior where $ \Re f > \varepsilon $, while inside the horizon cap, it vanishes.\footnote{Practically, we may take $\mathbb{1}_{h}$ as an analytical function, perform a desired computation on the contour and then at the end take the limit $\Lambda \to \infty$ followed by $\varepsilon \to 0$. The latter limit is to be understood as limit on the choice of contours where the horizon cap progressively approaches an infinitesimal circle around $r=r_{h}$.}

The slicing constraint now applies to $\tau$ and $s$ individually, giving  $\tau \in (0,1)$, $|s| \in (0,1)$.  Evaluating $\bsig$, we get
\begin{equation}
	\bbsig = \tau \ , \qquad \isig = s \, \mathscr{N}_{d} \ , \qquad \mathscr{N}_{d} = \frac{4^{\idim-1}\Gamma(\idim)^{2}}{\pi \Gamma(2 \idim)} \ ,
\end{equation}
where $\idim= \frac{1}{d}$. In writing the above, we have used the result
\begin{equation}
	\begin{split}
		\totr{\left(f^{\frac{1}{2}}\right)}  &=    \frac{f^{\frac{1}{2}}}{\pi i}   \, {}_{2}F_{1}\left(1- \idim,\frac{1}{2},\frac{3}{2},f \right) - i  \mathscr{N}_{d}    \ .
	\end{split}
\end{equation}
The takeaway is that, with a  judicial choice of $\sig$ such as in \cref{eq:SigInter0}, we can saturate the bound in \cref{eq:ReSigBound}.
The constant $\mathscr{N}_{d}$ controls the range of $\isig$ achievable using $\sig$ of the form \cref{eq:SigInter0} where $s$ is a constant. It is a monotonically increasing function of $d$, with  $\mathscr{N}_{2}=0.5$, $\mathscr{N}_{5} \approx 0.998$ and $\mathscr{N}_{d>5} > 1 $. We could have as well considered a more complicated form for $\sig_{\dif}$ than in \cref{eq:SigInter0} to further increase $\isig$. For example, one could promote $\tau$ and $s$ in \cref{eq:SigInter0} to functions of $r$ themselves, and optimize $\isig$ over profiles of $s(r)$. It can be argued that $s(r)$ of the form
\begin{equation}
s(r) = \frac{\tanh(\tilde{\Lambda}\sqrt{f})}{\sqrt{f}} \ , \quad \tilde{\Lambda} >> 1
\end{equation}
satisfies the slicing constraint and yields an unbounded value for $\isig$. Therefore we do not ascribe any special meaning to the numerical constant $\mathscr{N}_{d}$.

We may summarize this section as following. Allowable $\csig$ profiles fall under $3$
distinct classes based on the analytic properties of their respective tortoise transformations, which in turn fix the values of $\bsig$.   When expressed in terms of  the $\alpha$ coordinate, this corresponds to their parity structure under $\alpha \to -\alpha$. We summarize these properties in \cref{tab:AnProp}.
\begin{table}[h!] 
\begin{center}
	\begin{tabular}{|c | c|c|c|}
	\hline
$\csig(r): \csig(\alpha)$ & $\csig$(horizon) & $i \beta \totr{\csig}(r) :  i \beta \totr{\csig}(\alpha)$ & $\bsig$ \\
\hline
analytic : real even &  $\csig_{h} \neq 0$ & non-analytic : $i\beta \csig_{h}  \ctor(\alpha) +$ real even & $\csig_{h},$ real \\
analytic : real even &  $\csig_{h} = 0$ & analytic : real even & $0$ \\
non-analytic : real odd & $\csig_{h} = 0$ & non-analytic :  $i\beta \totr{\csig}(0) + $ real odd &  $ 2\totr{\csig}(0),$ imaginary \\
\hline
\end{tabular}
\end{center}
\caption{Analyticity and parity properties of bulk functions respectively in $r$ and $\alpha$ coordinates are given. In some cases, the reality properties are best expressed after shifting the function by a constant or a  piece proportional to $\ctor(\alpha)$. }
\label{tab:AnProp} 
\end{table}

\paragraph{Comment on terminology : } In the following text, we will address the radial contour in terms of the coordinate $\alpha$, $\ctor$ or simply $r$ as per convenience and hope this will not cause any confusion. In particular, we will qualify bulk profiles of $\sig, \csig$ and $\totr{\csig}$ as analytic/non-analytic and even/odd.

\subsection{Time reversal isometry }
The metric \cref{eq:sigmet} on the SKEW geometry  has a $\mathbb{Z}_{2}$ involution isometry given by
\begin{equation}
 \dd w \to -\dd w +   \frac{2 \sig}{r^{2} f} \dd r   = - \dd w + i \beta \, \dd \totr{\sig} \ ,
\end{equation}
originating from the time-reversal isometry of the Schwarzchild solution. This isometry acts on scalar field configurations as
\begin{equation}
 \Phi(w,r,\bx) \mapsto \Phi(-w+i\beta  \totr{\sig},r,\bx) \ . 
\end{equation}
As in analysis of the grSK case, the time reversal action can be used to relate solutions of field equations on the SKEW geometry\cite{Chakrabarty:2019aeu}. Introducing the Fourier space fields via 
\begin{equation}
 \Phi(w,r,\bx) = \int_{k} e^{ikx}\Phi(\w,r,\bk) \ , \qquad e^{ikx} := e^{-i\w w+i \bk.\bx} \ , \qquad \int_{k} := \int \frac{\dd \w}{2 \pi}\frac{\dd^{d} k}{(2\pi)^{d}} \ ,
\end{equation}
the time reversal action on the frequency space fields becomes
\begin{equation}
 \Phi(\w, r, \bk) \mapsto e^{-\beta \w \totr{\sig}}\Phi(-\w, r, \bk) \ .
\end{equation}
Following \cite{Jana:2020vyx}, we define a time-reversal covariant derivative 
\begin{equation}
 \bD_{\sig} := r^{2}f\del_{r} + \sig \del_{w} \   \xrightarrow{\text{Fourier trans.}}  r^{2}f\del_{r} - i \sig \w \ .
\end{equation}
We use $\bD_{\sig}$ to represent both the Fourier and real domain derivatives, and hope the meaning will be clear from the context.  $\bD_{\sig}$ satisfies
\begin{equation}
\begin{split}
 \bD_{\sig}\Phi(w,r,\bx) =&\ \int_{k} e^{ikx}(r^{2}f\del_{r}-i \sig \w)\Phi(\w,r,\bk) \ , \\
 \bD_{\sig}\Phi(-w+i\beta \totr{\sig},r,\bx) =&\ \int_{k} e^{ikx}e^{- \beta \w \totr{\sig}} 
\  (r^{2}f\del_{r}+i \sig \w)\Phi(-\w,r,\bk)  \ .\\
 \end{split}
\end{equation}
We also use the notation
$
\bD_{\pm} := \bD_{\pm 1} \ .
$
Further, $\bD_{\sig}$ satisfies 
\begin{equation} \label{eq:shiftsig}
 e^{-\beta \w  \, \totr{\kappa}} \bD_{\sig} \left(e^{\beta \w \, \totr{\kappa}}  \Phi\right) =  \bD_{\sig+2\kappa} \left( \Phi\right)  \ ,  \qquad     e^{-\beta \w \totr{\csig}}\bD_{\sig}\left( e^{\beta \w\totr{\csig}} \Phi\right) = \bD_{+}( \Phi) \ .
\end{equation}
The above relation will turn out handy in next section when studying field perturbations on SKEW geometries.  
 
\subsection{Probing with scalar field} \label{sec:Probes}

We now study a probe scalar field in the SKEW geometry. Though it is straightforward to work with the general class of designer scalars introduced in \cite{Ghosh:2020lel}, as an illustrative example, let us work with an ordinary massless Klein-Gordon scalar field. The following discussion can easily be extended to all designer scalar probes and hence to gauge and metric perturbations (i.e., Markovian and non-Markovian probes in \cite{Ghosh:2020lel}). The scalar field action is given by
\begin{equation} \label{eq:KGAct}
	S= -\frac{1}{2}\int \dd x^{d+1} \sqrt{-g} \, \nabla^{A}\Phi \nabla_{A}\Phi + \frac{1}{2(d-2)}\int\dd x^{d} \sqrt{-\gamma} \, \gamma^{\mu \nu} \partial_{\mu}\Phi \partial_{\nu}\Phi \ ,
\end{equation}
where $g$ $(\gamma)$ denotes bulk (boundary) metric and upper Latin (lower Greek) letters denote corresponding indices (see \cref{fn:BdySign}). The boundary counterterms (up to second order in derivatives) \cite{Ghosh:2020lel} have been included in the above. The equation of motion of $\Phi$ is 
\begin{equation}\label{eq:KGeqn}
 r^{1-d}\bD_{\sig}\left(r^{d-1}\bD_{\sig} \Phi\right) +(\w^{2}-f k^{2})\Phi = 0 \ 
\end{equation}
in terms of frequency space variables ($\bD_{\sig}=r^{2}f\del_{r}-i\sig\w$).
To obtain a regular solution to  \cref{eq:KGeqn} near  future (or past) horizon, we need to set $\sig(r_h) = 1$ (or $-1$) as in Eddington-Finkelstein coordinates.\footnote{For a single sided geometry this condition implies that we are passing on to the Eddington-Finkelstein  gauge very close to the horizon, where $\sig$ should satisfy an appropriate smoothness condition. But since we wish to analytically continue the metric on to radial contour, $\sig$ is taken to be analytic in a finite domain around the contour.} To study the solutions to \cref{eq:KGeqn}, it is convenient to anchor our discussion in reference to such a nice coordinate system. Therefore, we express the field perturbation for the background with a general $\sig$ in terms on the same `physical' perturbation in ingoing gauge.  This is achieved by defining
\begin{equation}\label{eq:IntoSigDiff}
 \Phi(\w,r,\bk) = e^{\beta \w\totr{\csig}}   \varphi(\w,r,\bk) \ ,
\end{equation}
 where $\varphi$ denotes the field solutions in ingoing coordinates and solves
\begin{equation}\label{eq:KGingo}
 r^{1-d}\bD_{+}\left(r^{d-1}\bD_{+}\varphi \right) + (\w^{2}-fk^{2})\varphi = 0 \ .
\end{equation}
The dressing with $e^{\beta \w \totr{\csig}}$ implements the diffeomorphism which relates the present coordinates to ingoing coordinate system for frequency space fields.\footnote{{For a more general tensor field, this gauge transformation also involves contractions with the flow matrix $\flow\index{^\beta_\delta}$ defined in \cref{eq:SigFlow}.}} To derive \cref{eq:KGingo} from \cref{eq:KGeqn}, we have used the relation in \cref{eq:shiftsig}. The solutions of the above ODE which are regular at the horizon describe ingoing (quasi-normal) modes. According to the classification of designer scalars, $\Phi$ is a Markovian scalar probe, implying that the natural boundary condition on it is of Dirichlet type \cite{Ghosh:2020lel}, which we normalize by demanding $ \lim_{r \to \infty} \varphi^{\text{in}} =1 $. 

\subsubsection{SK Solution and dual operators} 

The  parameterization of field perturbations given by \cref{eq:IntoSigDiff} allows us to directly import various results from the literature for quasi-normal modes to  SKEW spacetime. We use the results and conventions in \cite{Ghosh:2020lel} as needed to express the ingoing solution of \cref{eq:KGingo} and its boundary dispersion kernel.

The general solution on the SKEW contour is
\begin{equation}\label{eq:SKsol}
	\Phi_{\sSK} (\w, \ctor, \bk)=\text{exp}\left( \beta \w \totr{\csig}\right) \left[ J_{i} \varphi^{\text{in}}(\w,\ctor,\bk) + J_{h} e^{-\beta \w \ctor} \varphi^{\text{in}}(-\w,\ctor,\bk) \right] \ ,
\end{equation}
where $J_{i}$ and $J_{h}$ respectively parameterise ingoing solutions and their out-going counterparts obtained through the action of time-reversal symmetry. Imposing boundary conditions
\begin{equation}
\lim_{r \to \infty + i 0} \Phi_{\sSK} = J_{\sL} \ ,  \qquad  \lim_{r \to \infty - i 0} \Phi_{\sSK} = J_{\sR} \ , 
\end{equation}
implies
\begin{equation}
	J_{i} + J_{h} = J_{\sL} \ , \qquad J_{i} + e^{-\beta \omega}J_{h} = e^{-\beta \omega \bsig} J_{\sR}\ ,
\end{equation}
giving
\begin{equation}\label{eq:PFvar}
	J_{i} = (\nb+1) e^{-\beta \w \bsig}J_{\sR} - \nb J_{\sL}  \ , \qquad J_{h} = -(\nb+1) \left(e^{-\beta \w \bsig}J_{\sR} -  J_{\sL} \right) \ ,
\end{equation}
where $\bsig$ (defined in \cref{eq:bsig}) and $\nb$ are (see \cref{fn:nBRel})
\begin{equation}
\bsig = \lim_{r \to \infty - i 0} \totr{\csig} \ , \qquad 	\nb = \frac{1}{ e^{\beta \omega}-1}\ .
\end{equation}
Thus the combination of boundary sources \cref{eq:PFvar}  is uniquely selected out by monodromies of $\totr{\csig}$ and $\ctor$ (along the SKEW contour) across the asymptotic boundaries. The solution defined by substituting \cref{eq:PFvar} in \cref{eq:SKsol} matches with the modular transformed two sided black hole solution identified in \cite{Son:2009vu}.

We now proceed to the evaluation of the on-shell action with the solution in \cref{eq:SKsol}. First we note that the normal derivative of $\Phi$ at a constant radius hypersurface is given by
\begin{equation}
	\del_{n}\Phi \equiv n^{A}\del_{A} \Phi  = \frac{1}{r |\sqrt{f}|} \bD_{\sig} \Phi \ ,
\end{equation}
where $n^{A}$ denotes the outward normal to the constant $r$ hypersurface (at the boundary $r\to \infty - i  0$).\footnote{A potential ambiguity in the sign of $n^{A}$ arises at the $\sL$ boundary ($r\to \infty + i  0$) as the factor $\sqrt{f}$ itself changes sign. But this is only an issue of semantics. One can factor in the right sign either into $\sqrt{f}$ factors or work with $|\sqrt{f}|$ and absorb the sign into the orientation of $n^{A}$. We follow the latter convention. Similarly, our choice of counterterms are in terms of $|\sqrt{f}|$, which we explicitly \textit{subtract} from the $\sL$ boundary. Again, both approaches are consistent as boundary counterterms always appear with $\sqrt{f}$ factors from $\sqrt{\gamma}$.\label{fn:BdySign}} The (renormalized) conjugate radial momentum density of $\Phi$ is
\begin{equation}
	\begin{split}
		\Pi &= -   \sqrt{-\gamma} \,  \del_{n} \Phi + \text{counterterms} \\
		&=  - \text{exp}\left(\beta \omega \totr{\csig}\right)\left[ r^{d-1} \bD_{+}   +  \frac{r^{d-1} }{d-1} |\sqrt{f}|  \left( \partial_{i} \partial_{i} - f^{-1} \partial_{w}^{2} \right) \right] \varphi \ ,
	\end{split}
\end{equation}
where the latter term in the second line (accompanied by $\sqrt{f}$ factor) is the contribution from counterterms. $\Pi$ gives the renormalized radial momentum of $\Phi$ when evaluated at the respective boundaries.  An analytic renormalized radial momentum density can be defined for $\varphi^{\text{in}}$ by stripping away the factor $e^{\beta \omega \totr{\csig}}$. In expression,
\begin{equation}
	\begin{split}
		K^{\text{in}}(\omega,\vb{k}) & := - \lim_{r \to \infty} e^{-\beta \omega \totr{\csig}} \Pi \big|_{\varphi = \varphi^{\text{in}}(\w,\vb{k})}  \\
		& = b^{-d}\left\{- i b \omega - \frac{(bk)^{2}}{d-2} + \ldots  \right\} \ .
	\end{split}
\end{equation}
Here we have  extracted $K^{\text{in}}$ from the ingoing solution expressed in the $\sig =1 $ gauge of SKEW geometry or the original grSK geometry \cite{Ghosh:2020lel}. Let us define the frequency reversed version of $K^{\text{in}}$ as
\begin{equation}
K^{\text{rev}}(\omega,\vb{k}) := K^{\text{in}}(-\omega,\vb{k}) \ .
\end{equation}
The dual boundary operators are 
\begin{equation}
	\langle \mathcal{O}_{\sL/\sR}\rangle = \lim_{r \to \infty \pm i 0} \Pi \ .
\end{equation}
Evaluating the above for $\Phi_{_{\text{SK}}}$ gives
\begin{equation}\label{eq:Mod1pt}
\begin{split}
	\langle \mathcal{O}_{\sL}(w,\vb{k}) \rangle  &= -K^{\text{in}}(w,\vb{k})\left[(\nb+1)e^{-\beta \omega \bsig}J_{\sR}-\nb J_{\sL} \right] + (\nb+1) K^{\text{rev}}(\omega,\vb{k})\left[ e^{-\beta \omega \bsig}J_{\sR}-J_{\sL}\right] \\
	\langle \mathcal{O}_{\sR}(w,\vb{k}) \rangle  &= -K^{\text{in}}(w,\vb{k})\left[(\nb+1)J_{\sR}-\nb e^{\beta \omega \bsig}J_{\sL} \right] + \nb K^{\text{rev}}(\omega,\vb{k})\left[ J_{\sR}-e^{\beta \omega \bsig}J_{\sL}\right] \ .
\end{split}
\end{equation}

\subsubsection{Generalized Influence Functional}

We now evaluate the on-shell action with the solution in \cref{eq:SKsol}.  The corresponding boundary action is termed the \emph{generalized influence functional} (GIF), and is the one parameter generalization of Feynman-Vernon influence functional \cite{Feynman:1963fq} capturing the effect of modular flow.  

Since we consider the bulk theory to only quadratic order in fields, the on-shell action reduces to boundary terms linearly depending on the boundary radial momentum. In the present case, we generate two such terms, one from each boundary. The resultant GIF for the probe $\Phi$ is 
\begin{equation}\label{eq:GIF}
	\begin{split}
		S_{\text{GIF}}[J_{\sR},J_{\sL};\bsig] &= \frac{1}{2} \lim_{r_{c} \to \infty} \int_{k} \left[ \Phi^{\dag} \Pi \right]^{r_{c} - i 0} _{r_{c}+i0}  \\
		 &= -\frac{1}{2} \int_{k}  \Big[ \left(J_{\sR}^{\dag} e^{\beta \w \bsig}- J_{\sL}^{\dag} \right) K^{\text{in}} \Big((\nb+1)J_{\sR}e^{-\beta \w \bsig} - \nb J_{\sL}  \Big) \\
		 & \quad \qquad \qquad - \left(\nb J_{\sR}^{\dag}e^{\beta \w \bsig} - (\nb+1) J_{\sL}^{\dag} \right) K^{\text{rev}} \Big( J_{\sR}e^{-\beta \w \bsig} -  J_{\sL} \Big)\Big] \\
		  &=- \int_{k}  \left(J_{\sR}^{\dag} e^{\beta \w \bsig}-J_{\sL}^{\dag} \right) K^{\text{in}} \Big((\nb+1)J_{\sR} e^{-\beta \w \bsig} - \nb  J_{\sL}  \Big) \\  
		  &=- \int_{k}  J_{h}(-\w, -\vb{k}) \frac{K^{\text{in}}(\w,\vb{k})}{\nb} J_{i}(\w, \vb{k}) \ ,\\ 
	\end{split}
\end{equation}
where in the last line we expressed the result in terms of $J_{i/h}$ combinations identified in \cref{eq:PFvar}. The above expression matches with the Son-Teaney result \cite{Son:2009vu} for modular flowed boundary action. It can be argued that the generalized analytical continuation considered by those authors to fix the field solution on the two-sided black hole was tailored to achieve this. But our treatment is apt for a rigorous understanding of the dynamics including interactions, and hence higher point boundary correlators. Before further comments on higher point correlations, let us record the $2$-pt correlations encoded in \cref{eq:GIF}. Varying \cref{eq:GIF}  (or \cref{eq:Mod1pt}) with corresponding sources yields two point correlations 
\begin{equation}\label{eq:Mod2ptBulk}
	\begin{split}
		\langle \mathsf{O}_{\sR}(\w, \vb{k}) \mathsf{O}_{\sR}(-\w, -\vb{k}) \rangle &= i K^{\text{in}}(\w,\vb{k}) + \nb \left( i K^{\text{in}}(\w, \vb{k})-i K^{\text{rev}}(\w, \vb{k})  \right) \ , \\ 
		\langle \mathsf{O}_{\sR}(\w, \vb{k}) \mathsf{O}_{\sL}(-\w, -\vb{k}) \rangle &=  \nb e^{\bsig \beta \w } \left( i K^{\text{in}}(\w, \vb{k})-i K^{\text{rev}}(\w, \vb{k})  \right) \ , \\ 
		\langle \mathsf{O}_{\sL}(\w, \vb{k}) \mathsf{O}_{\sR}(-\w, -\vb{k}) \rangle &= (\nb+1) e^{-\bsig \beta \w} \left( i K^{\text{in}}(\w, \vb{k})-i K^{\text{rev}}(\w, \vb{k})  \right) \ , \\ 
		\langle \mathsf{O}_{\sL}(\w, \vb{k}) \mathsf{O}_{\sL}(-\w, -\vb{k}) \rangle &= -i K^{\text{in}}(\w,\vb{k}) + (\nb+1) \left( i K^{\text{in}}(\w, \vb{k})-i K^{\text{rev}}(\w, \vb{k})  \right) \ .\\ 
	\end{split}
\end{equation}
Comparing with \cref{eq:ModKMSRel} and setting
\begin{equation}
	G^{_\text{Ret}}(\w, \vb{k}) = i K^{\text{in}}(\w,\vb{k}) \ ,
\end{equation}
we find agreement, proving that grSK contour produces the expected modular flowed correlations.\footnote{We could also define $K^{\text{rev}}(\w,\vb{k})$ with a momentum reversal, making the relation $K^{\text{rev}}(\w,\vb{k}) = K^{\text{in}}(\w,\vb{k})^{*}$ more accurate. However, in the present system which preserves parity this does not change any result.}

\subsubsection{Comments on higher point functions} \label{sec:GIFKMS}

It is straight forward to extend our results to generate modular transformed higher point correlation functions of the boundary. In the gravitational description, such correlations correspond to scattering processes on the SKEW geometry, which may in turn be computed using Feynman-Witten diagrams. Though we do not endeavor to study such processes in this work, motivated from previous results for the grSK geometry, we can immediately make interesting inferences about their properties. 

Notice that the final form of the qudratic order GIF reported in \cref{eq:GIF} is at most linear in both the combinations $J_{i}$ and $J_{h}$. We expect this structure to generalize to all orders in perturbation theory as follows.  Consider the contribution to GIF which is $n$-th order in boundary sources, denoted as $S_{\text{GIF}}^{(n)}$.  Since $J_{i}$ and $J_{h}$ form two independent linear combinations of the boundary sources, $S_{\text{GIF}}^{(n)}$ can be written as an $n$-th order functional in $J_{i/h}$. We conjecture that this piece which generates connected $n$-pt correlators will be devoid of terms which are  $n$-th order exclusively in $J_{i}$ or $J_{h}$. In expression,
\begin{equation}
	S_{\text{GIF}}^{(n)} \big|_{J_{i}=0} =  S_{\text{GIF}}^{(n)} \big|_{J_{h}=0} =  0 \ .
\end{equation}
This statement can be translated to a condition on correlators using a dual basis of operators.  In terms of
\begin{equation}
\mathsf{O}_{i}(\w) := (\nb+1)e^{-\beta \w \bsig}\mathsf{O}_{\sR}(\w)-\nb \mathsf{O}_{\sL}(\w)   \qquad  \mathsf{O}_{h}(\w) :=  e^{-\beta \w \bsig } \mathsf{O}_{\sR}(\w) - \mathsf{O}_{\sL}(\w) \ ,
\end{equation}
we have that correlators made exclusively of $\mathsf{O}_{i}$ or $\mathsf{O}_{h}$ vanish. This is a direct generalization of the result first established in \cite{Chaudhuri:2018ymp} from microscopic unitarity and KMS relations for the case $\bsig=0$. It was then argued for contact interactions in holography using the grSK contour in \cite{Chakrabarty:2019aeu,Jana:2020vyx}  and subsequently verified for exchange interactions \cite{Loganayagam:2022zmq,Martin:2024mdm} (see also \cite{Loganayagam:2024mnj}).

Following the argument in \cite{Chakrabarty:2019aeu}, it is easy to validate our conjecture for contact Feynman-Witten diagrams. As an illustrative example, we turn on a cubic interaction in the action \cref{eq:KGAct} given by
\begin{equation}
S^{(3)}_{\text{int}} = \frac{\lambda}{3!}\int \dd x^{d+1} \ \sqrt{-g} \,  \Phi^{3}  + \text{boundary counterterms}\ .
\end{equation}
The contact contribution to the $3$-pt correlation is then obtained by integrating the triple product of boundary to bulk propagator along the SKEW contour. Alternatively, the leading correction to the boundary GIF in $\lambda$ is given by simply evaluating the interaction term using the SK solution in \cref{eq:SKsol}. In expression,
\begin{equation}
	\begin{split}
	S_{\text{GIF}}^{(3)} &= \frac{\lambda}{3 !} \int \displaylimits_{\ctor} \! \! \int \dd V_{\text{freq}}  \, \sqrt{-g} \prod_{\ell=1}^{3} \Phi_{\sSK} (\w_{\ell}, \ctor, \bk_{\ell})  \\
	&= \frac{\lambda}{3 !} \int \displaylimits_{\ctor} \! \! \int \dd V_{\text{freq}}  \, \sqrt{-g}  \\
	& \qquad \times \prod_{\ell=1}^{3} e^{ \beta \w_{\ell} \totr{\csig}} \left[ J_{i}(\w_{\ell}, \vb{k}_{\ell}) \varphi^{\text{in}}(\w_{\ell},\ctor,\bk_{\ell}) + J_{h}(\w_{\ell}, \vb{k}_{\ell}) e^{-\beta \w_{\ell} \ctor} \varphi^{\text{in}}(-\w_{\ell},\ctor,\bk_{\ell}) \right] \ ,
	\end{split}
\end{equation}
where we defined 
\begin{equation}
\int	\dd V_{\text{freq}} := \int \left(\prod_{\ell=1}^{3} \frac{\dd k_{\ell}^{d}}{(2 \pi)^{d}} \right) \, (2 \pi)^{d}\delta^{d} \left( \sum_{\ell=1}^{3} k_{\ell} \right) \ , \qquad \int \displaylimits_{\ctor}  := \ointctrclockwise \displaylimits_{\text{SKEW}} \! \dd \ctor \ .
\end{equation}
The integration over SKEW contour entails only the radial integration of the bulk $3$-pt contact vertex. $V_{\text{freq}}$ is defined including the momentum conserving delta function. Now let us analyze the following terms in $S_{\text{GIF}}^{(3)}$ which are cubic in $J_{i}$ and $J_{h}$:
\begin{equation}
\begin{split}
	\left(S_{\text{GIF}}^{(3)}\right)_{iii} &= \frac{\lambda}{3 !} \int \displaylimits_{\ctor} \! \!  \int \dd V_{\text{freq}}  \, \sqrt{-g} \prod_{\ell=1}^{3} e^{ \beta \w_{\ell} \totr{\csig}}   \varphi^{\text{in}}(\w_{\ell},\ctor,\bk_{\ell})  J_{i}(\w_{\ell}, \vb{k}_{\ell}) \ , \\
	\left(S_{\text{GIF}}^{(3)}\right)_{hhh} &= \frac{\lambda}{3 !} \int \displaylimits_{\ctor} \! \!  \int \dd V_{\text{freq}}  \, \sqrt{-g} \prod_{\ell=1}^{3} e^{ \beta \w_{\ell} ( \totr{\csig} - \ctor )}   \varphi^{\text{in}}(-\w_{\ell},\ctor,\bk_{\ell})  J_{h}(\w_{\ell}, \vb{k}_{\ell}) \ .
\end{split}
\end{equation}
Each factor in the integrands, other than the exponential factors $e^{- \beta \w \totr{\csig}}$ and  $e^{ \beta \w (\totr{\csig}-\ctor)}$ are analytic functions on the SKEW contour. Discontinuity in the integrands across the cut originates solely from these factors. In other words, if these factors were absent, the integrals should vanish -- the contributions from above and below the branch cut would cancel each other. The momentum conserving delta functions can be seen to exactly arrange this. The sum of frequencies from the three boundary to bulk propagators is constrained to vanish canceling out the exponential factors and in turn causing  the integrals to vanish. This proves the conjecture for $3$-pt correlations and immediately generalizes to contact diagrams involving arbitrary number of external legs.

For $(n > 3)$-pt correlators, one has to also verify this conjecture at the level of exchange diagrams. We expect it to hold, as it has been checked more thoroughly in the limit $\sig=1$ or the original grSK geometry. An additional subtlety occurs in cases with derivative weighted interactions which typically exist for interacting gauge bosons (and the Nambu-Goto string). In the grSK case, such interactions have been shown to generate specific localized contributions to $S^{(n)}_{\text{GIF}}$ from the horizon region on the radial contour \cite{Chakrabarty:2019aeu,Loganayagam:2022zmq}. Since our formalism allows for $\sig$ with non-trivial radial dependence, it is important to verify our conjecture explicitly in such cases.

An intuitive way to motivate our result is to notice that  SK generating functional with non-trivial $\bsig$  can be obtained from that with $\bsig=0$ by redefining the sources via
\begin{equation}\label{eq:JrSigTrans}
	J_{\sR}(\w) \to e^{-\beta \w \bsig } J_{\sR}(\w) \ ,  \qquad J_{\sL}(\w) \to J_{\sL}(\w)  \ .
\end{equation}
This transformation corresponds to translating $J_{\sR}$ sources along Euclidean time relative to that of $J_{\sL}$ sources.  Equivalently, the $\mathsf{O}_{\sR}$ operators are translated in the positive imaginary time directions relative to $\mathsf{O}_{\sL}$ via
\begin{equation}
	\mathsf{O}_{\sR}  \to e^{- \beta \bsig \mathsf{H}} \,  \mathsf{O}_{\sR} \, e^{ \beta \bsig \mathsf{H}} \ , \qquad \mathsf{O}_{\sL} \to \mathsf{O}_{\sL}  \ .
\end{equation}
This mirrors the structure of \cref{eq:ModNpt} after cycling through the $\rho^{\bsig}$ factor around the $\tr$ (see \cref{fig:SKTimeCon}).

While our prescription using the SKEW contour effectively reduces to \cref{eq:JrSigTrans} at the level of $2$-pt correlations, its advantage shows when generalizing to the case of interactions and correspondingly higher point correlations in the boundary, as evidenced by the special case of grSK contour. As another special case of our formalism, we may obtain higher point TFD correlations for which an effective prescription has been lacking in the literature.\footnote{Shock wave scattering near the horizon of an eternal black hole  was famously used by \cite{Shenker:2014cwa} to estimate the OTO 4-pt correlator. However, we do not know how to systematically improve such computations from a straight forward bulk perturbation theory which includes effects from interactions supported on the full  geometry.}

\section{A closer look at the horizon}\label{sec:RindlerSKEW}

Let us now take a closer look at the analytical continuation in radial coordinate, focusing on its action near the horizon. We start by considering a finite cut out region $M$ near the  horizon of the original single sided geometry. The region $M$ is  approximately flat  and can be parameterized using a Gaussian normal coordinate system anchored on the horizon as,
\begin{equation}\label{eq:GNcoords}
	\dd s^{2}|_{M} \approx 4 \, \dd v^{+} \dd v^{-} + g_{_\perp}\dd \vb{y}^{2}  \ .
\end{equation}
\begin{figure}[h!]\label{fig:RindWedge}
	\begin{center}
		\includegraphics[scale=0.2]{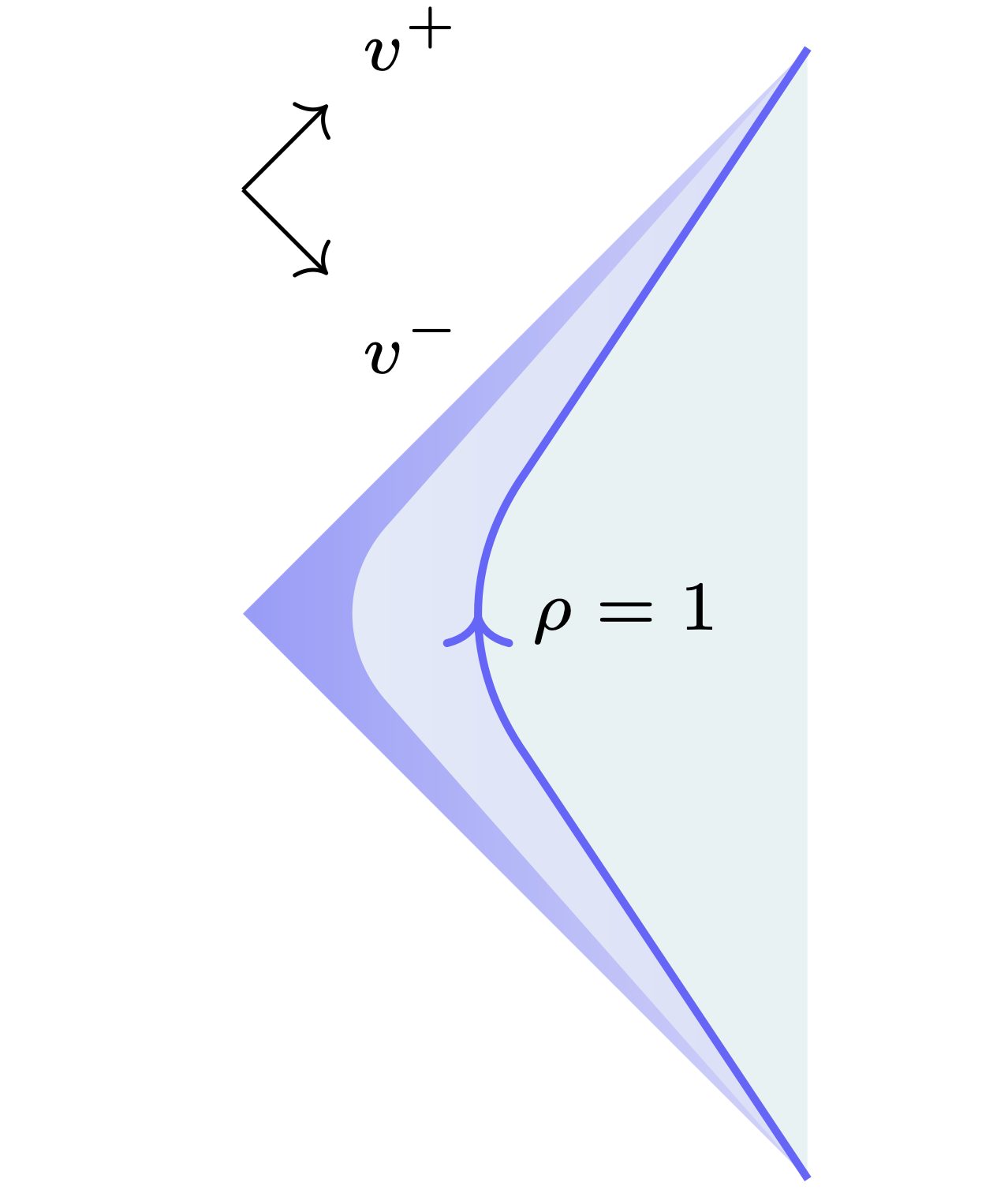}
	\end{center}
\caption{The near horizon region $M$ is shown in light blue with its boundary $\rho:=v^{+}v^{-}=1$ marked. The arrow shows flow of time $t$. The corresponding stretched horizon region which becomes part of the horizon cap is  shown in dark blue. More generally, $M$ denotes the spacetime close to an entanglement horizon.}
\end{figure}
Here $v^{\pm}$ denote the pair of null normal vectors to the horizon whose origins align with the past and future horizons respectively. The tangential directions to the horizon are labelled $\vb{y}$ and the corresponding metric $g_{_\perp}$ (any $v^{\pm}$ dependence therein is suppressed). Since, we are typically concerned with the exterior geometry, for now, we consider the patch $v^{\pm}>0$ and $v^{+}v^{-}<1$.  To make contact with our original coordinate system in \cref{eq:sigmet}, we introduce Rindler like coordinates $t \in \mathbb{R}, \rho \in (0,1)$ via
\begin{equation}\label{eq:NullCor}
v^{+} := \rho \, \text{exp} \left( \frac{t}{2 \ell_{T}} - \int_{\rho_{c}}^{\rho} \frac{\vsig(z)  \dd z}{z}\right) \ , \qquad  v^{-} := \rho \, \text{exp} \left( -\frac{t}{2 \ell_{T}} + \int_{\rho_{c}}^{\rho} \frac{\vsig(z)  \dd z}{z}\right) \ ,
\end{equation}
where $\ell_{T} $ sets a horizon or temperature scale via $ 4 \pi \ell_{T} = \beta$, and $\rho_{c}$ is some reference position which can as well be fixed as $\rho_{c}=1$. The precise domain of $\rho$ is not important for the present discussion, provided $M$ covers a finite region including the stretched horizon. Note that unless $\vsig$ is set to identically vanish, $t=0$ does not describe a surface of time-reflection symmetry in the above parameterization.
$\vsig(\rho)$ now plays the role of $\sig$  and satisfies $\vsig^{2}<1$ as in \cref{eq:Sigbound}. Consequently, the time slices we consider always pass through the point $\rho=0$ resulting in a coordinate singularity. Expressing the metric using \cref{eq:NullCor}, we get
\begin{equation}
	\dd s^{2} = -\frac{\rho^{2}}{\ell_{T}^{2}} \left( \dd t + \frac{2 \ell_{T}}{\rho} (1-\vsig) \dd \rho\right)  \left( \dd t  -\frac{2 \ell_{T}}{\rho} (1+\vsig) \dd \rho\right) + g_{_\perp}\dd \vb{y}^{2} \ ,
\end{equation}
which matches the  near horizon limit of \cref{eq:sigmet} (up to an overall scale), if we identify
\begin{equation}
w \sim t \ , \quad \sqrt{f} \sim \rho \ , \quad \sig \sim \vsig\ , \qquad r^{2}f' \sim \ell_{T}^{-1} \ , \qquad \frac{\dd r}{r^{2}f}  = \frac{2}{r^{2} f'}\frac{\dd \sqrt{f}}{\sqrt{f}} \sim 2 \ell_{T} \frac{\dd \rho}{\rho} \ ,
\end{equation}
where $f'=\frac{\dd f}{\dd r}$. To complexify $M$ to its SKEW version denoted as $\breve{M}$, we have to consider a radial contour defined via $\rho = \alpha - i 0$, and analytically continue the coordinates \cref{eq:NullCor} to $\alpha \in (-1,1)$.\footnote{The coordinate $\alpha$ in this section is only the near horizon limit of the $\alpha$ in \cref{eq:AlphaDef}. We hope the context resolves any potential ambiguity.}  The range of $\alpha$ has been fixed to be symmetric around the turning point $\alpha=0$. We may work directly with the variable $\rho$, provided we take to be slightly shifted towards the lower half of complex $\rho$ plane. An infinitesimal region around $\alpha=0$ now plays the role of an horizon cap which joins the two copies of $M$ with $\alpha > 0$ and $\alpha<0$, denoted $M_{\pm}$ respectively. 

Each point $p = (v^{+},v^{-}, \vb{y})$ in $M$ now has two images in $\breve{M}$, one in each of $M_{\pm}$, which we denote by $p_{\pm}$ respectively. Comparing the images across the horizon cap, we find
\begin{equation}\label{eq:RelBoost}
p =	(v^{+},v^{-},\vb{y})|_{p_{+}} =(v^{+} e^{2 \pi i \, \msig(p) },v^{-} e^{2 \pi i \, (1-\msig(p)) },\vb{y} )|_{p_{-}}  \ , \qquad \msig(p) :=\frac{1}{2 \pi i} \int\displaylimits_{-|\rho| }^{ |\rho|  } \frac{1 - \vsig(\alpha)}{\alpha - i 0} \dd \alpha
\end{equation}
where $|\rho| = \sqrt{\Re v^{+} v^{-}} $. The first equation in \cref{eq:RelBoost} is a diffeomorphism that maps regions $M_{\pm}$ to $M$ which acts trivially on $M_{+}$ while on $M_{-}$ it involves a local boost with a complex boost angle $\msig(p)$. We can think of \cref{eq:RelBoost} as the natural map which \textit{folds} the region $M_{-}$ back on to the original region $M$. Alternatively, the analytical continuation implemented by the SKEW contour is such that the two images of $p$ are relatively boosted with respect to each other with a complex rapidity factor $e^{2 \pi i \, \msig}$. The coordinate systems in $M_{\pm}$ are smoothly interpolated across the complexified horizon cap region to this effect.

We can classify profiles of $\vsig$ just like the classification of $\sig$ based on their monodromy properties on the contour. The analogue of $\sig$ profiles with trivial monodromy are now the $\vsig$ profiles which are even functions of $\alpha$. Similarly, $\sig$ profiles with a square root monodromy correspond to $\vsig$ profiles which are odd functions of $\alpha$. In particular, when $\vsig$ is an even function of $\alpha$, $\msig$  receives only a localized contribution from $\alpha=0$ and can be read out as
\begin{equation}
\vsig \text{  even:    } \qquad   \msig (p)  =\lim_{\alpha \to 0} \frac{1 - \vsig(\alpha)}{2} = \msig_{\text{total}}\ ,  
\end{equation}
where $\msig_{\text{total}}$ denotes the relative imaginary boost between $p_{+}$ and $p_{-}$ with $|\rho|=1$. The integral in \cref{eq:RelBoost} defining $\msig(p)$ is the counterpart of  tortoise transformation introduced in \cref{eq:tortrans} for the present case. The value of $\msig_{\text{total}}$ is the analogue of Euclidean modular time $\bsig$ discussed in \cref{sec:BBSKEW}. Evidently, for an even real function $\vsig$ the corresponding $\msig_{\text{total}}$ is also real.  As discussed in \cref{sec:SigProf}, it is possible to generate imaginary contribution to $\msig_{\text{total}}$ by considering profiles of $\vsig$ which are odd functions of $\alpha$.  For example, 
%
%\begin{equation}
%\frac{1 - \vsig(\alpha)}{2}= \tanh (\tilde{T}\alpha) , \qquad \msig(p) \approx \frac{2}{i \pi} \ln \left(\frac{4 e^{\gamma_{_E}} \tilde{T}}{\pi} |\rho | \right) , \qquad  \msig_{\text{total}}(\tilde{T}) =   \frac{2}{i \pi} \ln \left(\frac{4 e^{\gamma_{_E}} \tilde{T}}{\pi} \right),
%\end{equation}
%
%
\begin{equation}\label{eq:LocReBoost}
	\vsig =  \tanh (\tilde{T}\alpha) , \qquad  \msig_{\text{total}}(\tilde{T}) \approx   \frac{1}{2}+\frac{i}{ \pi} \ln \left(\frac{4 e^{\gamma_{_E}} \tilde{T}}{\pi} \right),
\end{equation}
where $\tilde{T}  \gg 1$ and $\gamma_{_E}$ is Euler's constant.\footnote{To independently generate $\Re \msig$ and $\Im \msig$ we can again use a horizon cap supported indicator function as in \cref{eq:SigInter0}.} Across the boundary $|\rho|=\pm 1$ of $M$, we can thus arrange for any amount of real modular parameter (or imaginary $\msig_{\text{total}}$) fixed in terms of $\tilde{T}$.  Thus, though $M$ describes only the near horizon region of the full geometry, $\Im \msig_{\text{total}}$ generated across it is unbounded.

The nature of the near horizon coordinate system explained above gives various interesting insights into the SKEW construction, which we now discuss.

\subsection{Relation to eternal double sided geometries }\label{sec:TwistTFD}

We begin by comparing the geometry of $\breve{M}$ with the double sided Kruskal like extension $M_{\text{et}}$ of $M$. Recall that $M_{\text{et}}$ is constructed by analytically continuing $v^{\pm}$ from the positive quadrant (the original $M$) to other quadrants on the $v^{\pm}$ plane. As is well understood, this geometry has a second copy of $M$ we denote $M'$ where $v^{\pm}<0$ (they respectively are the right and left wedges in the usual parlance). In addition, $M_{\text{et}}$ also includes a pair of future/past interior  regions.

A coordinate system which covers only the subregion $\widetilde{M} := M \cup M' \subset M_{\text{et}}$ can be obtained from \cref{eq:NullCor}  as follows. We set $\vsig$ to vanish identically and extend the coordinate domain as $\rho \in [-1,0) \cup (0,1] $. Alternatively, for every point $p = (v^{+},v^{-},y)\in M$, $\widetilde{M}$ also includes its reflection along the bifurcation point given by $p' = (-v^{+},-v^{-},y) \in M'$.\footnote{Usually, one uses two different radial co-ordinates for $M$ and $M'$, say $\rho_{\sR}$ and $\rho_{\sL}$, both positive. Any ambiguity is resolved if we use the light cone variables $v^{\pm}$ to make comparisons.} The slice $t=0$ is then invariant under the time reversal action $t \to -t $. On $M$, $\del_{t} v^{+}$ ($\del_{t} v^{-}$) is positive (negative) while on $M'$ it is negative (positive).  Also, $\del_{\rho} v^{\pm} > 0$ everywhere on $\widetilde{M}$. This is shows that our parameterization of $M'$ agrees with the property that directions of $\del_t$ on the boundaries of $M$ and $M'$ inside $\widetilde{M}$ are aligned opposite with respect to each other. The point $\rho=0$ has a coordinate singularity and hence is not included in the $(t,\rho)$ coordinate system. As the reader might have already noticed, the geometries $\widetilde{M}$ and  $\breve{M}$ (with $\vsig = 0$) are closely related but with an interesting twist.

Let us quickly recall the standard near horizon physics in an eternal geometry. Strictly speaking the bifurcation point $v^{+}=v^{-}=0$ does not belong to $\widetilde{M}$, making it the disjoint union of the sister universes $M$ and $M'$. In order to identify field modes on $\widetilde{M}$ which have regular extensions to $M_{\text{et}}$, we need to impose some physical boundary conditions to relate modes supported on $M$ and $M'$.  The usual trick to achieve this is to work with modes which satisfy appropriate boundary conditions close to the horizons $v^{\pm}=0$ ($t \to \pm \infty$) \cite{Birrell:1982ix}.  We refer to these boundary conditions as \textit{eternal TFD boundary conditions}. They demand that the ingoing (outgoing) excitations in $M$ locally have positive (negative) energy in $M_{\text{et}}$. Equivalently, one may demand that the correct field modes on $\widetilde{M}$ satisfy certain analyticity properties when compared across $M$ and $M'$ -- eternal TFD field modes are analytic in the lower $v^{\pm}$ complex plane. If we denote the ingoing and outgoing modes in $M$ as 
\begin{equation}\label{eq:Mmodes}
\psi_{\w}(|\rho|) e^{-i \w t} \sim (v^{+})^{-i\frac{\beta}{2 \pi} \w} \ ,\qquad  \psi_{\w}^{*}(|\rho|) e^{-i \w t} \sim (v^{-})^{ i \frac{\beta}{2 \pi} \w } , 
\end{equation}
respectively, the extended modes supported on full $\widetilde{M}$ selected out by the TFD boundary conditions are given by
\begin{equation}\label{eq:TFDmodes}
e^{-i \w t}  \left[  \psi_{\w}(|\rho|)\big|_{\sR} + e^{-\frac{\beta}{2} \w} \psi_{\w}(|\rho|)\big|_{\sL}  \right] \ , \qquad  e^{-i \w t}  \left[  \psi_{\w}^{*}(|\rho|)\big|_{\sR} + e^{\frac{\beta}{2} \w} \psi_{\w}^{*}(|\rho|)\big|_{\sL}  \right] \ .
\end{equation}
Here $\psi_{\w}(|\rho|)\big|_{\sL} =\psi_{\w}(|\rho|)\big|_{\sR} =\psi_{\w}(\rho)$ is an even function of $\rho$, which essentially follows from the $\rho \to -\rho $ symmetry of background metric.\footnote{This continues to hold for the full blackbrane geometry as field modes are functions of the radius $r$ which is in turn an even function of $\alpha$.} Notice that the same solutions as in $\cref{eq:TFDmodes}$ could have been identified if we  extended the modes in \cref{eq:Mmodes} to  $\widetilde{M}$ by continuing $\rho \to - \rho $ in \cref{eq:NullCor} along the lower half complex $\rho$ plane to implement the continuation from $v^{\pm}>0$ to $ v^{\pm}<0$.\footnote{In terms of $u:=-v^{-}$, the analyticity is along the upper half plane. Therefore the combined continuation on $v, u$ is equivalent to $t \to t - i \frac{\beta}{2} $.}

Now let us compare with the analogous procedure  on $\breve{M}$. The quantum fields on $\breve{M}$  meet no coordinate issues owing to the complexification of the near horizon region -- the coordinate singularity is avoided by the SKEW contour. In this case, analyticity in $\alpha$ uniquely picks out the field modes for us. Clearly, we obtain the same modes as in \cref{eq:TFDmodes}. Alternatively, as evidenced by the boundary correlators, both procedures pick out physically identical solutions on the corresponding geometry. Therefore it is tempting to ask if the two geometries can be identified (or distinguished) on the exterior regions away from the horizon cap.

\begin{figure}[h!]
	\begin{center}
		\includegraphics[scale=0.21]{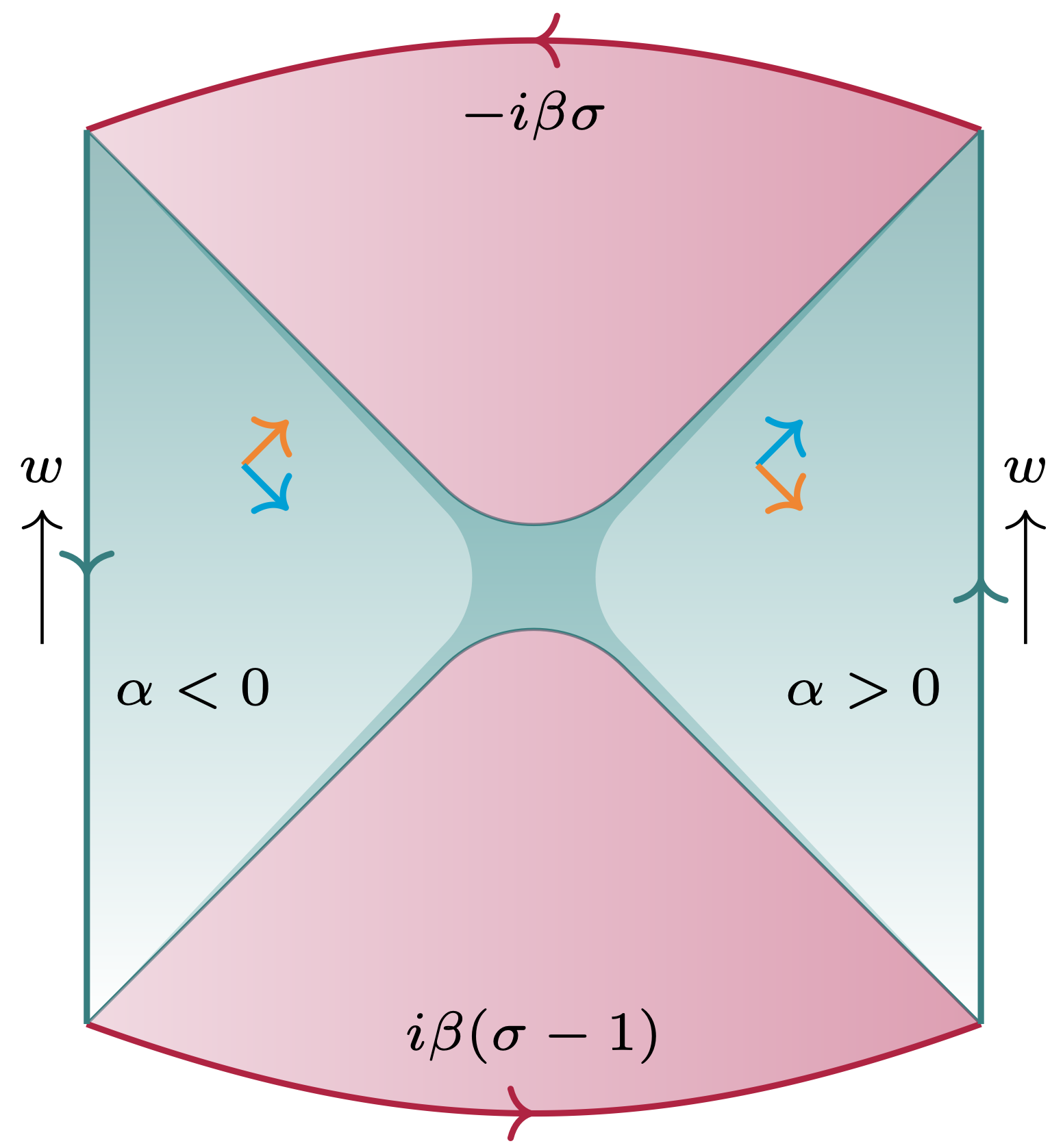} 
		\qquad   \qquad 
		\includegraphics[scale=0.22]{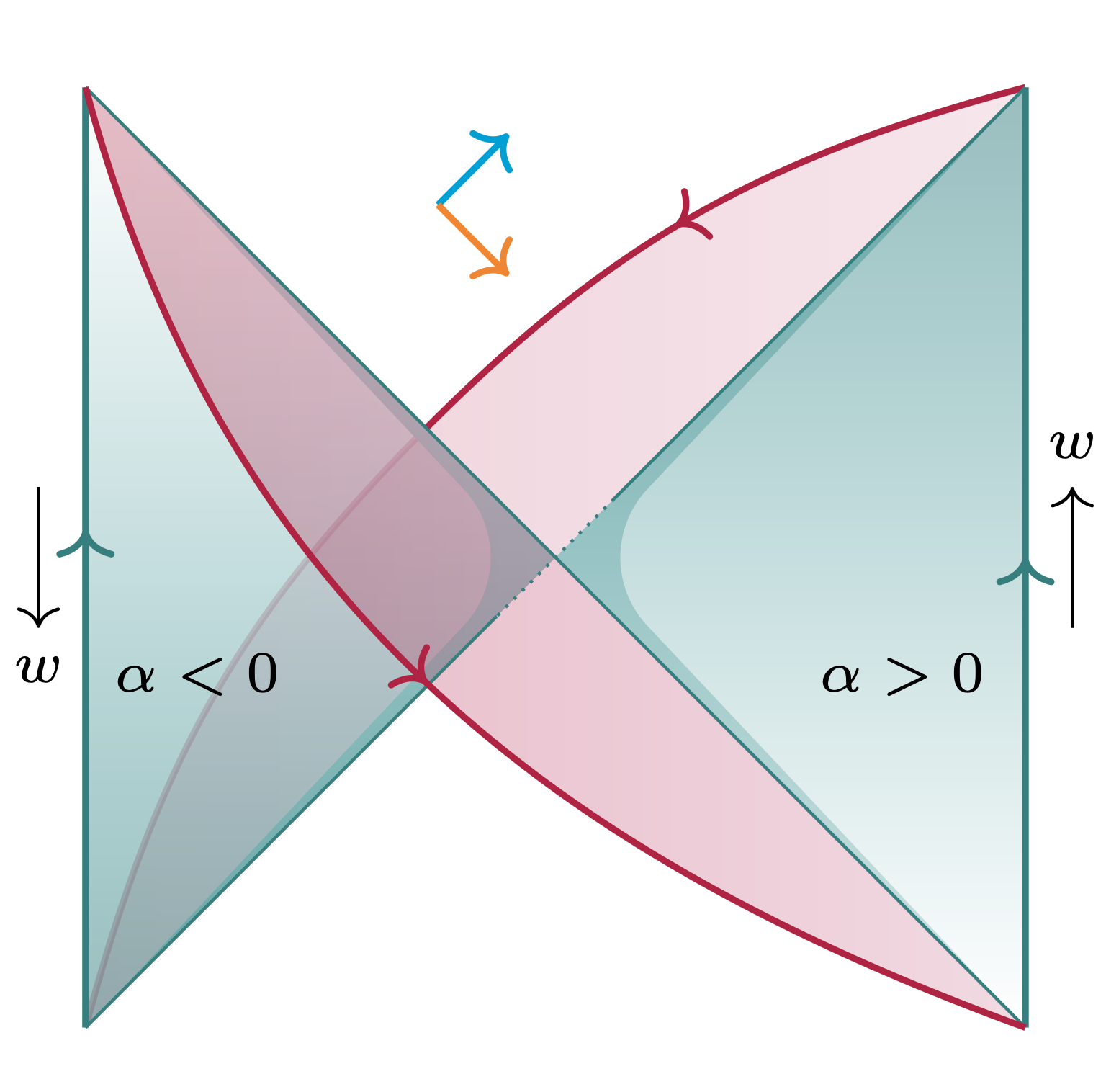}
	\end{center}\caption{Left: The schematic topology of the SKEW geometry is shown. Lorentzian region  covered by the radial contour is shown in teal, with complimentary  Euclidean regions (in purple) schematically attached to the past and future most time slices. The Euclidean caps are complementary sections of the cigar geometry. The SK contour time flow is marked in respective colours.   The horizon cap (shaded darker) smoothly interpolates between the two exterior regions. Right: The TFD limit ($\bsig=\frac{1}{2}$) of the same geometry is shown, with the Lorentzian region outside the stretched horizon embedded in the two sided eternal geometry. The Euclidean caps now form complementary half discs. The region inside the stretched horizon is excised out and replaced by the horizon cap. The region with $\alpha<0$ is embedded in the left wedge, and the reversal of Schwarzschild time direction gives the geometry a \textit{twisted ribbon} structure when embedded in the $\mathbb{C}^{2}$ of complex $v^{\pm}$ directions. The gloabl Kruskal time flow on the spacetime boundary matches the Schwinger-Keldysh contour time flow. In either figure, the red regions denote Euclidean subsections and should not be confused with the interior. The null axes for real $v^{\pm}$ (valid for the TFD configuration) are shown in cyan and orange respectively.}\label{fig:GGSKTFD}
\end{figure}

Denoting by $I$ the domain $(-1,1)$ of $\alpha$, and by $\mathbb{R}$ the domain of $t$,  $\breve{M}$ has the topology  $I \times \mathbb{R}$, albeit with a complexified metric on a smaller strip region in the middle that marks the horizon cap. As a compactified version, we may think of $\breve{M}$ as a finite strip. The time flow $\del_t$ on either boundaries $\alpha=\pm 1$ are aligned identically on this geometry, say upwards. Therefore moving a point upwards on $M_{-}$ corresponds to decreasing $v^{+}$ and increasing $v^{-}$. In addition, $\del_{\alpha}v^{\pm}$ continue to be positive.  This implies that the SKEW continuation flips the orientation of the $v^{\pm}$ frame across the horizon cap, while on $\widetilde{M}$ it was preserved by construction. Let us now consider a continuous map $\mathscr{F}: \breve{M} \to \widetilde{M}$ which respects the identification $M=M_{+}$ outside the stretched horizon. Such an $\mathscr{F}$ identifies the flow of $t$ at the boundaries of $M_{+}$  ($\alpha=1$)  and $M$ ($\rho=1$). In order to also align the flow of $t$ at the second boundary $\alpha=\rho=-1$, we conclude that, $\mathscr{F}$ should \textit{twist} $\breve{M}$ across the horizon cap into a ribbon like structure when viewed as an embedding in the $\mathbb{C}^{2}$ formed by complex $v^{\pm}$ coordinates. Consequently,  the initial and final time slices of $\breve{M}$
get mapped to the horizons $v^{+}=0$ and $v^{-}=0$ of $\widetilde{M}$ respectively (see  \cref{fig:GGSKTFD}).  In a strict sense, the horizons $v^{\pm}=0$ are not part of $\breve{M}$ as they are replaced by the horizon cap. As depicted in the left panel of \cref{fig:GGSKTFD}, in the conformal diagram of the geometry, the far future and past time slices mostly lie inside the horizon cap.

The above discussion shows that both eternal TFD boundary conditions and the SKEW analytical continuation at the TFD configuration $\vsig=0$ select out identical field modes, provided we identify the manifolds via the map $\mathscr{F}$ as described above. Extrapolating that result to the case of the full black brane implies that the SKEW geometry (with $\bsig = \frac{1}{2}$) can also be interpreted as describing the dynamics of an  entangleld TFD state of two distinct copies of the holographic CFT. From \cref{eq:TFDComm}, operators inserted on the two boundaries do commute with each other (yet their correlations do not factorize due to the entanglement in the state). In addition, we get a resolution of the coordinate singularity at the bifurcation point within a co-ordinate chart which covers only the exterior regions at the price of an \textit{emergent and complexified horizon cap}.  Again, the real advantage of having a resolved view into the bifurcation point manifests only when one studies higher point correlations. By studying  general choices of $\vsig$ we can also go away from the TFD limit and probe non-trivial modular evolved configurations of the geometry.

While so far we have been emphasizing the path integral perspective and in particular highlighting features of corresponding boundary correlators, we may also study quantum fields on these manifolds and their time evolution. We briefly comment on the time evolution operator in such geometries in \cref{sec:HorEn}. 

\subsubsection{The time evolution and horizon energy}\label{sec:HorEn}

In this section we work with $l_{T}=\frac{1}{2}$. The constant $t$ slice $\Sigma$ of $M$ has the unit normal  $ n^{\mu}$ and induced volume density $\sqrt{h}$. In expression,
\begin{equation}
n^{\mu}= \frac{1}{2}\left(\frac{ \upsilon}{\rho},  -\frac{\vsig }{\upsilon}, \vb{0}\right) \ ,\qquad \sqrt{h} = 2 \upsilon \sqrt{g_{_\perp}} \ , \qquad \upsilon:=\sqrt{1-\vsig^{2}} \ ,
\end{equation}
when evaluated for $\rho>0$. For quantum fields with stress tensor $T_{\mu \nu}$, the generator of $t$ flow is given by
\begin{equation}\label{eq:HorHam}
	\begin{split}
	H_{\vsig} &= 2 \int_{\Sigma}   \sqrt{g_{_\perp}}\upsilon \,  n^{\mu}  T_{\mu t}  : = \int_{\Sigma} \mathcal{H}_{\vsig}(\rho)\ , \qquad 
	 \mathcal{H}_{\vsig}(\rho) = \sqrt{g_{_\perp}} \left(\frac{\upsilon^{2}}{\rho} T_{tt} - \vsig T_{t\rho} \right)\ ,
	\end{split}
\end{equation}
where the integration measure is given by $ \dd \rho \dd \vb{y}$. 
The half sided Hamiltonian $H_{\vsig}$ can be extended to the double sided geometries $\breve{M}$ or $\widetilde{M}$. Depending on whether the SKEW extension $\breve{M}$ or usual Kruskal extension $\widetilde{M}$ is considered, we denote the time slices as $\breve{\Sigma}$ and $\widetilde{\Sigma}$ respectively. The difference between the two is essentially that $\widetilde{\Sigma}$ excludes the bifurcation locus $\rho=0$, and hence evaluates the time evolution on the disjoint union of intervals $[-1,0)\cup (0,1]$, while in $\breve{\Sigma}$  the time slice is connected via the horizon cap. Corresponding Hamiltonians are defined as $\breve{H}_{\vsig}$ and $\widetilde{H}_{\vsig}$ respectively. In expression,
\begin{equation}
	\begin{split}
		\widetilde{H}_{\vsig}&= \int\displaylimits_{\widetilde{\Sigma}}  \dd \rho \, \mathcal{H}_{\vsig}(\rho) \ , \qquad \breve{H}_{\vsig} = \int \displaylimits_{\breve{\Sigma}} \dd \alpha \,  \mathcal{H}_{\vsig}(\alpha-i0)  \ , \\
	\end{split}
\end{equation}
where we have suppressed integration on the transverse space with $\dd \vb{y}$.  Let us for a moment focus on the TFD configuration $\vsig=0$ ($\upsilon =1$), for which $\widetilde{\Sigma}$ can be identified with the time slice of the usual Rindler spacetime. In this case $\widetilde{H} := \widetilde{H}_{\vsig=0} $ decomposes as $\widetilde{H} = K + K'$, where
\begin{equation}
	K:= \int \displaylimits_{0}^{1}\frac{\dd \rho}{\rho}\, \sqrt{g_{_\perp}}T_{tt}\ , \qquad K':=   \int \displaylimits_{-1}^{0}\frac{\dd \rho}{\rho}\, \sqrt{g_{_\perp}}T_{tt} \ .
\end{equation}
The meaning of $K$ and $K'$ requires some clarification. Note that, for $\rho<0$, the correct surface normal aligned towards $\del_t$  is $-n^{\mu}$.  Therefore, $K$ evolves fields on $M$ along $t$ flow, while $K'$ evolves fields on $M'$ opposite to the direction of $t$ flow.  This corresponds to time evolution along the SK contour at the  boundary.\footnote{$\widetilde{H}$ implements the global Kruskal time evolution usually denoted as $H_{\sR}+H_{\sL}$ in the literature and is \textit{not} an isometry of the double sided TFD state. The operator which implements $t$ evolution on both $M$ and $M'$ is given by $K-K' \sim H_{\sR}-H_{\sL}$, which is an isometry of the geometry. \label{fn:GlobalHam}}  This can be compared with the corresponding expression for $\breve{H}:=\breve{H}_{\vsig=0}$,
\begin{equation}
	\breve{H}= \int \displaylimits_{-1}^{1} \frac{\dd \alpha}{\alpha- i 0}\, \sqrt{g_{_\perp}}T_{tt} = \widetilde{H} + Q,\ .
\end{equation}
where $Q$ is the residue contribution (if any) from the horizon. $\breve{H}$, just like $\widetilde{H}$, evolves fields supported on $M_{+}$ forward in $t$ and fields supported on $M_{-}$ backward in $t$. In addition, it also involves the contribution denoted as $Q$ which acts on field supported on the horizon cap. In this sense $Q$ captures the part of `energy' that resides in the horizon cap.  We can try and extend the notion of $Q$ to all SKEW geometries via 
\begin{equation}
	\breve{H}_{\vsig} = \widetilde{H}_{\vsig} + Q_{\vsig} \ ,
\end{equation}
where $\widetilde{H}_{\vsig}$ is the Hamiltonian computed without the horizon cap (read principal value contribution). Further, in a gravitational theory, the matter stress tensor sources the metric and the energy content of a spacetime point is encoded in the local geometry. Therefore we may evaluate $Q_{\vsig}$ directly from the near horizon properties of the geometry. In expression,
\begin{equation}\label{eq:GeoHorEn}
 Q_{\vsig}=\frac{i}{4 G_{N}} \int \displaylimits_{\vb{y}} \Res\displaylimits_{\rho=0} \left(\upsilon \sqrt{g_{_\perp}} n^{\mu }\delta \mathcal{G}_{\mu t} \right) \ ,
\end{equation}
where we have reinstated the transverse integral and $\delta \mathcal{G}_{\mu \nu}$ is the variation in the Einstein tensor (together with the cosmological term) due to back reaction from $T_{\mu \nu}$.\footnote{More generally, it is the difference of the equation of motion operator evaluated on the geometry back reacted by $T_{\mu \nu}$ and the unperturbed SKEW background. If $T_{\mu \nu}$ is defined to include gravitons of quadratic order and onwards, this reduces to linear graviton terms.} We have isolated its strength at $\rho=0$ using the $\Res\displaylimits_{\rho=0}$ symbol. The relation \cref{eq:GeoHorEn} suggests that we should identify the evolution operator localized to the horizon cap as some geometric operator acting on the horizon.

The definition of $Q_{\vsig}$ can be generalized straightforwardly to the full SKEW contour of the black brane, but its essence is already captured from the near horizon analysis.  Notice that $Q_{\vsig}$ is the Hamiltonian version of the statement recorded in \cref{sec:GIFKMS} that, certain kinds of interactions (more precisely, derivative coupled interactions) produce extra localized contributions from the horizon cap to the boundary influence functional.  We interpret it as capturing a regulated version of IR fluctuations in the CFT which manifest in the bulk as near horizon fluctuations. We will not attempt a detailed analysis of what kind of interactions and bulk states lead to such contributions in this work.

We conclude this section with the following comment. We can extend the operators $\breve{H}_{\vsig}$ and $Q_{\vsig}$ to the full SKEW geometry, which we define as $\breve{H}_{\sig}$ and $Q_{\sig}$ respectively. Since $\breve{H}_{\sig}$ implements time translation along the boundary SK contour (see \cref{fn:GlobalHam}), its action on bulk fields should be equivalent to the diffeomorphism generated by $\sqrt{f}\delta \csig_{\dif}$ which shifts $\isig = \Im \bsig$. Therefore we have,
\begin{equation}
	i\del_{\isig} \propto \breve{H}_{\sig} \ ,
\end{equation}
as an operator on the bulk Hilbert space of semi-classical quantum fields on the SKEW geometry. Note that the stress tensor density $T_{tt}$ in \cref{eq:HorHam} is to be constructed out of perturbative matter fields and gravitons excited on the SKEW geometry. We expect $\breve{H}_{\sig}$ to have a finite expectation value even when we take $G_{N}$ to be small. To ensure that $\breve{H}_{\sig}$ is a genuine operator on the Hilbert space, we need to further verify that it has bounded standard deviation in the small $G_{N}$ limit \cite{Witten:2021unn}. We hope to fill up these gaps in a future analysis.\footnote{Evolution operators on similar radial contours were also discussed in \cite{Chen:2023hra}.}

 \subsection{Connection to the JLMS relation}

The relative boost between the two images $M_{\pm} \subset \breve{M}$ implemented by the SKEW radial contour is closely related to the identification of bulk and boundary modular transformations argued in \cite{Jafferis:2015del, Jafferis:2014lza}. We will now quickly review this result which is applicable generically in AdS/CFT and point out its connection to the SKEW geometry.

Consider a generic AdS/CFT system with a smooth geometric dual. Let $\mathcal{R}$ be a connected spatial subregion of the boundary, $\bar{\mathcal{R}}$ its complement and $\Gamma$ be the corresponding quantum extremal surface in the bulk. $\Gamma$ gives a partition of the bulk Cauchy slice into two complementary regions $\mathsf{r}$ and $\bar{\mathsf{r}}$, where the boundary $\del \mathsf{r}$ of the former is given by $ \Gamma \cup \mathcal{R}$. Their respective entanglement wedges are denoted as $M_{\mathcal{R}}$ and $M_{\bar{\mathcal{R}}}$.

The  bulk surface $\Gamma$ is identified by the rules of quantum extremal surfaces \cite{Ryu:2006ef, Hubeny:2007xt} and has been studied in a variety of scenarios (see also \cite{Rangamani:2016dms}). Two results relevant to the present discussion are as follows. First, when the bulk geometry contains a black hole and $\mathcal{R}$ is sufficiently large, $\Gamma$ approaches the black hole horizon.\footnote{In \cref{sec:BBSKEW} we focused on the planar limit and $\mathcal{R}$ was the full spatial slice of the boundary and satisfies this.} Second, for empty bulk geometries and ball shaped boundary subregions, one can via conformal transformation of the boundary, map the problem to that of a topological black hole as elaborated in \cite{Casini:2011kv}. The takeaway is that in either cases, the future entanglement horizon (or boundary of $M_{\mathcal{R}}$) can be treated similar to the acceleration  horizon of a Rindler observer, which allows a comparison with the near horizon analysis of the SKEW geometry. The observables supported on $\mathcal{R}$ depend on the reduced density operator on $\mathcal{R}$, which we specify without loss of generality using the modular Hamiltonian $K_{\mathcal{R}}$. We can move away from stationary configurations by turning on boundary sources (and include their back reaction perturbatively). This corresponds to adding a few particles in the bulk and such a state falls into the category of $U(1)$ non-invariant states discussed in  \cite{Lewkowycz:2013nqa, Faulkner:2013ana}. The state is assumed to give a smooth geometry for $M_{\mathcal{R}}$.

According to \cite{Jafferis:2015del}, $K_{\mathcal{R}}$ can be identified with the modular Hamiltonian $K_{\mathsf{r}}$ of bulk fields supported on $\mathsf{r}$, up to a term proportional to the area operator localized on $\Gamma$. In expression,
\begin{equation}\label{eq:JLMS}
K_{\mathcal{R}} = \frac{A_{\Gamma} }{4 G_{N}} + K_{\mathsf{r}} + \ldots  \ ,
\end{equation}
where $A_{\Gamma} $ denotes the area operator and $\ldots$ denotes corrections subleading in $G_{N}$. The explanation for the combination of terms on the right hand side of \cref{eq:JLMS} roughly goes as follows. The bulk modular Hamiltonian $K_{\mathsf{r}}$ when extended into $\bar{\mathsf{r}}$ such that  $K_{\text{bulk}} = K_{\mathsf{r}} -K_{\bar{\mathsf{r}}}$ has a regular action everywhere on $M_{\mathcal{R}}$ and $M_{\bar{\mathcal{R}}}$. Alternatively, the $K_{\text{bulk}}$ is an outer automorphism of the algebras of observables supported on $M_{\mathcal{R}}$ and $M_{\bar{\mathcal{R}}}$.  Though in general, $K_{\text{bulk}}$ has a non-local action within $M_{\mathcal{R}}$, very close to the entanglement horizon, one can appeal to the local flat physics to deduce that $K_{\text{bulk}}$ acts identical to the local boost operator whose acceleration horizon coincides with the entanglement horizon of $M_{\mathcal{R}}$ \cite{Bisognano:1976za,Hislop:1981uh}. However, it is only the half sided version $K_{\mathsf{r}}$ of $K_{\text{bulk}}$ which appears in \cref{eq:JLMS}. The action of $K_{\mathsf{r}}$ is not regular, as it is supported only on $M_{\mathcal{R}}$ and produces a `kink' at $\Gamma$ \cite{Jafferis:2014lza}. The combination with the area operator $A_{\Gamma}$ is designed precisely to compensate for this singular behaviour such that $K_{\mathcal{R}}$ on the left hand side of \cref{eq:JLMS} has a regular action in the bulk. A corollary is that the relative modular operator between a pair of states has a regular action without the area term \cite{Jafferis:2015del}.

Let us now explain how the above discussion relates to the SKEW prescription.  To study the real time dynamics of the reduced state on  $\mathcal{R}$, we can adopt the SKEW formalism as follows. 
We first set up a Gaussian normal coordinate system anchored on $\Gamma$ as in \cref{eq:GNcoords}.  A  SKEW geometry corresponding to $M_{\mathcal{R}}$ can be constructed by parameterizing the null coordinates as in \cref{eq:NullCor} and analytically continuing the radial direction. The resulting geometry $\breve{M}_{\mathcal{R}}$ has two images of $M_{\mathcal{R}}$ we denote as $M_{\mathcal{R}\pm}$,  conjoined via a slightly thickened and complexified version of the entangling surface $\Gamma$ (which is analogous to the horizon cap).
By construction, the coordinate system on $\breve{M}_{\mathcal{R}}$  will be such that points in $M_{\mathcal{R}\pm}$ are boosted with respect to each other as in \cref{eq:RelBoost}. This relative boost is inherited by all field modes supported on $M_{\mathcal{R}\pm}$, which we interpret as the universal part of the half sided modular evolution, but now on the SK description of the dynamics of $M_{\mathcal{R}}$ on $\breve{M}_{\mathcal{R}}$ instead of its natural purification living on $M_{\mathcal{R}} \cup M_{\bar{\mathcal{R}}}$.

As noted before, in the JLMS relation \cref{eq:JLMS}, the area operator $A_{\Gamma}$ serves to regulate the singular action of the half sided boost close to $\Gamma$. In contrast, the SKEW description is smooth by construction as we have regulated the bifurcation point by slightly deforming the spacetime into the complex radial (or lapse) direction. As a price we are forced to include an additional contribution to the evolution operator $\breve{H}_{\vsig}$ given by $Q_{\vsig}$. Further, in \cref{eq:GeoHorEn}, we motivated that $Q_{\vsig}$ should be thought of as a geometric operator localized on the bulk entangling surface (or horizon), albeit being a purely imaginary quantity.  In this sense, $Q_{\vsig}$ plays a role analogous to the area operator $A_{\Gamma}$ in  \cref{eq:JLMS}. We do not have a deeper understanding of this observation at the moment and would like to investigate it in a future work.

\subsection{Modular flowed geodesics}

Let us now highlight a closely related property implied by the near horizon structure of the SKEW geometry. We wish to argue that the SKEW construction motivates a modified geodesic approximation for modular flowed $2$-pt functions. We will first describe this problem, conjecture a solution inspired from the SKEW construction and discuss why our conjecture is closely related to the  prescription of \cite{Faulkner:2018faa}.\footnote{I thank Onkar Parrikar for bringing this work to my attention.}

 We again consider the situation where the boundary is divided into complementary subregions $\mathcal{R}$ and $\bar{\mathcal{R}}$, but now focus on boundary $2$-pt correlators of some heavy operator $O_{\Delta}$. The dominant contribution to this correlation function can be extracted via the geodesic approximation as $\langle O_{\Delta}(y_{1}) O_{\Delta}(y_{2}) \rangle \sim e^{-\Delta\ell_{12}}$, where $\ell_{12}$ is the bulk geodesic distance between the boundary points $y_{1,2}  \in \del M_{\mathcal{R}}$.
We now ask how should one generalize the geodesic approximation to compute modular flowed correlations of the form 
\begin{equation}
\langle e^{-2\pi\bsig K_{\mathcal{R}}} O_{\Delta}(y_{1}) e^{2\pi\bsig K_{\mathcal{R}}}O_{\Delta}(y_{2}) \rangle =:  e^{-\Delta \ell_{12}^{\bsig} } \ .
\end{equation}
For large $\Delta$ one expects that there exists some $1$-dimensional bulk curve $\frak{c}$ which is anchored to the boundary and whose length approximates $\ell_{12}^{\bsig}$ (after appropriate regularization). 
We propose to identify $\frak{c}$ with the geodesic on the SKEW geometry (with appropriate choice of $\vsig$) which is anchored to  points $y_{1}$ and $y_{2}$ on the boundaries $\del M_{\mathcal{R}+}$ and $\del M_{\mathcal{R}-} $ respectively. In other words, instead of working with the original spacetime region $\mathcal{R}$ we interpret the correct $\frak{c}$ as a curve in the corresponding SKEW geometry.
Such a geodesic has to necessarily pass through the horizon cap region which in turn shrinks to an infinitesimal complexified region close to $\Gamma$ as one considers radial contours arbitrarily close to the real axis.

Now we can take a step back, and ask if there is a curve which can be defined within the original patch $M_{\mathcal{R}}$ that computes the quantity $\ell^{\bsig}_{12}$. To be precise, we consider the case $\Re \vsig = \frac{1}{2}$. For the local Rindler patch, we can use the map  \cref{eq:RelBoost} to obtain an image of $\frak{c}$ within the region $M_{\mathcal{R}}$. In principle, we can extend the relation \cref{eq:RelBoost} to the full SKEW geometry, which gives a map from $\breve{M}_{\mathcal{R}}$ to $M_{\mathcal{R}}$ which \textit{folds} the region $M_{\mathcal{R}_{-}}$ back on to $M_{\mathcal{R}}$. The folded image of $\frak{c}$ on $M_{\mathcal{R}}$ which we term $\tilde{\frak{c}}$, consists of a pair of geodesics $\tilde{\frak{c}}_{1}$ and $\tilde{\frak{c}}_{2}$ attached to the boundary at $y_{1}$ and the image of $y_{2}$ (under the extension of \cref{eq:RelBoost}) respectively. In the limit where horizon cap is taken to be small, both $\tilde{\frak{c}}_{1}$ and $\tilde{\frak{c}}_{2}$ are anchored to $\Gamma$ in the bulk. It is convenient to think of $\tilde{\frak{c}}$ as a single geodesic which starts from the boundary and reflects off $\Gamma$ back to the boundary.  Since the metric is regular near  $\Gamma$, one does not expect the length functional evaluated on $\frak{c}$ to receive significant contributions from the horizon cap. Therefore one should be able to estimate $\ell^{\bsig}_{12}$  from the single sided curve $\tilde{\frak{c}}$.

Since we already established the effectiveness of SKEW geometries in computing correlations of bulk fields, the geodesic prescription can be thought of as the leading order WKB approximation for the relevant correlation function. It is natural to ask if we could have directly identified the curve $\tilde{\frak{c}}$ without going through the SKEW construction.  This problem was studied by the authors of  \cite{Faulkner:2018faa} using a modified geodesic prescription motivated from the JLMS relation and the replica trick. While they analyzed more general scenarios involving correlations across complementary boundary sub regions, our interest is in the case involving `mirror operators' in their terminology \footnote{This terminology is originally from \cite{Papadodimas:2013jku} where the mirror operators excite field modes behind the horizon.} which corresponds to the case $\Re \bsig =\frac{1}{2}$. The result of \cite{Faulkner:2018faa} is to fix $\tilde{\frak{c}}_{1,2}$ such that they are attached to the corresponding boundary points $y_{1,2}$ on $\del M_{\mathcal{R}}$, but attached to the entangling surface $\Gamma$ with a relative (real) boost. The amount of boost is fixed according to the modular evolution desired in the boundary correlator. Recall that the basic feature of \cref{eq:RelBoost}, and its generalizations to the full SKEW geometry is also that they implement a non-vanishing relative boost between the subregions across horizon cap. Therefore our heuristic description of $\tilde{\frak{c}}$ is qualitatively similar to that of  \cite{Faulkner:2018faa}.  It would be interesting to study our prescription for the modular flowed geodesic correlators in more detail in a future work.

\section{Generalizations of the bulk state}\label{sec:ModAvgStates}

 In \cref{sec:Probes} we showed how to set up the analysis of probe scalar fields in general SKEW backgrounds. While this analysis can be upgraded to include perturbative backreactions on the geometry, the scope of our analysis is limited to near-equilibrium scenarios as the dynamics we discuss describe semi-classical deformations of the black brane sourced from the boundary. In this section we comment on two generalizations of the bulk state.

\subsection{Excitations above the thermal state}\label{sec:ExThermal}

Recall that in the computation of the GIF in \cref{eq:GIF}, we had ignored any boundary terms from the future and past time slices of the geometry. The tacit assumption there was that such contributions do not contribute to the influence functional. One plausible argument to justify this assumption is that, as we have hinted before in  \cref{fig:GGSKbulk,fig:GGSKTFD}, the SKEW geometry can be completed using appropriate sections of the Euclidean cigar geometry in the future and past following \cite{Skenderis:2008dg, Skenderis:2008dh}. Ignoring boundary terms in the on-shell action is thus equivalent to the assumption that there are no non-trivial boundary sources on the Euclidean segments.

However, completing the geometry with such Euclidean segments also raises various conceptual issues. Firstly, we do not know how to smoothly attach such Euclidean sections to SKEW geometries while also respecting the analytical structure of the horizon cap. A cut and glue construction would naively conflict with the advantages of a smooth bulk. Further, the analytical structure of the higher order interactions we discussed in \cref{sec:GIFKMS} was based on the assumption that bulk interactions are supported on the Lorentzian sections. It is not clear how this structure, which as is already leads to the correct KMS properties, modifies (or stays preserved) if we were to allow for interactions supported on the Euclidean extensions.

Therefore we suggest an alternative standpoint, where one terminates the bulk geometry at the future and past infinite time slices of the SKEW geometry without any additional Euclidean segments. We briefly discuss this possibility now. In this case, one can argue away the contributions to bulk on-shell action from the time slices by adding appropriate boundary terms to these slices as follows. We first regulate the SKEW geometry by restricting it to the time window
$(-W,W)$. Further, we extend the probe action in \cref{eq:KGAct} to include the terms
\begin{equation}\label{eq:TimeBdyAct}
S_{\text{time slice}} = \frac{1}{2} \int \dd \alpha \, \dd x^{d-1}   \Big[\sqrt{\nu} \,  m^{A} \Phi \del_{A}\Phi +  \ldots \Big]_{w=-W}^{w=W} \ ,
\end{equation}
where $m^{A}$  and $\nu$ denote the unit normal and induced metric on the time slice respectively. The boundary terms \cref{eq:TimeBdyAct} are designed to cancel the boundary contributions generated from the bulk. The limit of large $W$ can be taken subsequently. While this prescription does not particularly clarify the variational principle on the time slices, it justifies our computation of the GIF. It follows that modifying the boundary terms in \cref{eq:TimeBdyAct} would give rise to additional contributions to the GIF. In semi-classical perturbation theory we may interpret this as modifying the bulk state by injecting (collecting) particles from the past (future).

Our prescription to fix the boundary data on a time slice that extends into the bulk may naively seem to violate diffeomorphism invariance on these slices and hence unacceptable in a gravitational theory. However, in the large $W$ limit, both the future and past time slices approach the respective horizons, which are diffeomorphism invariant surfaces (see \cref{fig:GGSKTFD}). For sufficiently small number of particles insertions, we expect that the horizon can be used as a reference surface with respect to which semi-classical excitations can be defined. The excited states we have in mind should be thought of as gravitationally dressed to the horizons (see \cite{Bahiru:2022oas,Bahiru:2023zlc} for a discussion of bulk operators dressed to features of the background state).  We hope to report a more elaborate analysis of such excited states on SKEW geometries in a future work.

\subsection{SK ensemble observables } \label{sec:ClassSigEns}

If we were to parameterize the SKEW metric in an ADM like decomposition, $\bsig$ can be related to the $0$-mode of the radial component of the shift vector (see \cref{fn:RadialADM}). In \cref{sec:SigProf} we analyzed possible ways to generate different values of $\bsig = \bbsig + i \isig $ adhering to the slicing constraint \cref{eq:Sigbound} and discovered that its real part $\bbsig$ is generated as a localized contribution from the horizon, while the imaginary part $\isig$ depends on the value of $\csig$ everywhere. In this sense $\bbsig$ should be thought of as an \textit{edge mode contribution} from the horizon and $\isig$ is a genuine \textit{$0$-mode} on the SKEW geometry. Recall that the combinations of boundary sources appearing in \cref{eq:PFvar} only depend on the total monodromy of $\totr{\csig}$ across the boundaries and not per se on how the monodromy is distributed in the bulk. 

This observation clarifies the nature of diffeomorphism invariance on the SKEW geometry. Any allowable deformations $\delta \totr{\csig}$ of the bulk time slices should be such that they do not act on the asymptotic boundaries, or equivalently $\delta \totr{\csig}$ should itself be an analytic function and  satisfy $\lim_{\alpha \to \pm1} \delta \totr{\csig}(\alpha) = 0$. It follows that they do not modify $\bsig$ and in turn the boundary source combinations \cref{eq:PFvar}.
They are generated by analytic deformations $\delta \csig_{\av}$ of the $\csig$ profile which are not supported on the horizon (see \cref{tab:AnProp}). Consequently, the combination $i \beta \delta \totr{\csig}_{\av}$  need not vanish at the horizon, but will take real values everywhere on the geometry including the horizon, and correspond to deformations of the bulk time slices via \cref{eq:DiffToIn} within the same SKEW geometry. The takeaway is that, in order to localize the boundary path integral on to a thermal contour with a fixed $\bbsig$ we need to also hold fixed the edge value of $\csig_{\av}$ at the horizon. Variations of $\csig_{a}$ supported on the horizon act as Euclidean diffeomorphisms on the boundary real time contour, which is reflected in the choice of solution picked out by the SKEW geometry. However, non-analytic deformations  $\sqrt{f} \delta \csig_{\dif}$ can generate different values for the $0$-mode $\isig$ even when supported away from the horizon. In fact, in a field theory, the collection of correlation functions one can generate from any thermal contour with a given value of $\bbsig = \Re \bsig$ and infinite span of real time between its Euclidean segments will be indistinguishable. The choice of $\isig$ only relabels the operator insertions by translating the ket operators relatively with respect to the bra operators along the real time direction.

Nevertheless, as discussed in \cref{sec:Intro}, semi-classical observables in the bulk form a Type II von Neumann algebra following the crossed product construction in \cite{Witten:2021unn}. The corresponding Hilbert space consists of semi-classical excitations over modular averaged TFD states \cref{eq:CPstate} of complementary exterior regions, with boundary observables \cref{eq:CPObs}.  We seek a counterpart of these observables in the open EFT of the exterior, which as before we define as the dynamics on the ingoing configuration of the SKEW geometry (or alternatively, the grSK saddle). To maintain that the open EFT description of semi-classical dynamics is complete (or equivalent to the TFD description), we need to distinguish between SKEW geometries with unequal values of $\isig$. Moreover, we need to allow classical ensembles of such geometries to be part of the open EFT.

A physical justification for this suggestion is that any practical protocol followed by an exterior observer to assign a coordinate system on the background geometry is bound to have inherent fluctuations, which translates to working with an ensemble of gauge choices. Suppose that, instead of choosing a fixed gauge as in \cref{eq:sigmet}, we conceptualize the background geometry as an average over close lying gauge choices. This is similar to the Faddeev-Popov method. But here we are not referring to the gauge fixing procedure for perturbative gravitons above the background, but rather for the graviton condensate which forms the black hole geometry itself. As evidenced by \cref{eq:Mod2ptBulk}, single sided observables dual to boundary correlators of $\sR$ (or $\sL$) operators do not distinguish between different gauge choices in this ensemble. But once analytically continued to the SKEW saddle, some elements of this gauge ensemble will clearly correspond to non-analytical deviations in the $\csig$ profile and correspond to boundary contours with different values of $\isig$. In this sense, promoting ensembles of SKEW backgrounds as genuine bulk states is equivalent to averaging over a distribution of gauges for the exterior geometry.

The above discussion may naively seem to suggest that one should average boundary observables over a small window of $\bbsig$ as well. However, in our formalism, the CGL type doubling of the exterior is carried out only after a bulk coordinate system is established. We interpret this as simply mimicking the field theory procedure, where identification and denomination of the relevant degrees of freedom logically precede their SK doubling. For an exterior observer the horizon marks a `boundary' of the observable universe, the SK ensemble is an average over only local deformations in $\csig \sim \del_{r} \totr{\csig}$ of the exterior away from the horizon.\footnote{In a regular QFT, the SK doubling does not promote freedom of field redefinitions in the single copy theory to new classical degrees of freedom. It is interesting to ask if there is an analogue of this phenomenon for ordinary gauge theories on the SK contour.}

Let us again specialize to the case of black brane solutions and the ingoing SK configuration of the SKEW geometry. Given that graviton perturbations are dual to the hydrodynamic modes of the open EFT in this case, a natural question is, how should one incorporate the ensembles of $\isig$ into the fluid gravity correspondence. We do not yet know a satisfactory answer to this problem. But we are tempted to conclude from the previous discussion that the \textit{possible states of the boundary fluid should also include classical moduli corresponding to $\isig$}. The moduli captured by $\isig$ are to be thought of as  classical degrees of freedom as there are no separate $\sR$ and $\sL$ versions thereof. The extended generating functional which encodes \emph{SK ensemble observables} \cref{eq:CPObsSK} of the open EFT is characterized by the temperature $\beta$ and the probability distribution $\mathsf{P}(\isig)$. In expression,
\begin{equation}\label{eq:ZCPinSK}
	\mathcal{Z}_{\beta, \mathsf{P}} [J_{\sR},J_{\sL}, x ] : = \int \dd \isig  \, \mathsf{P}(\isig) e^{i x \isig}  \,\tr \left\{ \left(\rho_{_{\beta}} \right)^{\bsig} \mathsf{U}[J_{\sR}]  \, \left(\rho_{_{\beta}} \right)^{1-\bsig} \, \mathsf{U}^{\dagger}[J_{\sL}] \right\}\Big|_{\bbsig = 0} \ ,
\end{equation}
where $x$ is a source for $\isig$ and $J_{\sL/\sR}$ are the SK sources for other fields in the system. The observables defined in \cref{eq:CPObsSK} can be generated by acting on $\mathcal{Z}_{\beta, \mathsf{P}}$ with $\pm\frac{\delta}{i\delta J_{\sR/\sL}}$ and $X(\frac{\del}{i \del x })$, and subsequently setting the sources to vanish. Clearly, there is a straightforward generalization  of \cref{eq:ZCPinSK} to other values of $\bbsig$ and in particular to the outgoing SK frame where $\bbsig = 1$. 

Let us conclude with the following remark. We do not yet have a bulk argument that fixes the allowable choices for $\mathsf{P}(\isig)$. However, we might need to impose some physical restrictions on $\mathsf{P}(\isig)$. For example, we demand that allowable $\mathsf{P}(\isig)$ satisfy the Hermiticity condition 
\begin{equation}\label{eq:EvenP}
\mathsf{P}(\isig) = \mathsf{P}(-\isig) \ .
\end{equation}
The reason is that observables in open EFT include expectations of Hermitian operators of the form
\begin{equation}\label{eq:CompHermOp}
\langle \mathcal{O}^{\dag}\mathcal{O} \rangle_{\rho_{_\beta}}\ , \qquad \mathcal{O} := \mathbb{T} \left\{ \mathsf{O}_{1}\mathsf{O}_{2} \ldots \right\} \ .
\end{equation}
The corresponding expectation value generated from an SK ensemble state is given by
\begin{equation}\label{eq:OdO}
\langle  \mathcal{O}^{\dag} \mathcal{O}\rangle_{\rho_{_\beta}, \mathsf{P}} = \int \dd \isig \, \mathsf{P}(\isig) \, \text{tr} \left\{ \left(\rho_{_{\beta}} \right)^{i \isig} \mathcal{O} \left(\rho_{_{\beta}} \right)^{1-i \isig} \mathcal{O}^{\dag} \right\} \ ,
\end{equation}
which continues to be real for $\mathsf{P}(\isig)$ satisfying \cref{eq:EvenP}. Notice that \cref{eq:EvenP} is insufficient to recover the positive definiteness of $\langle \mathcal{O}^{\dag}\mathcal{O} \rangle_{\rho_{_\beta}}$.

\section{Discussion}

The primary objective of this work was to extend the  CGL prescription for the gravitational SK saddle \cite{Glorioso:2018mmw} to construct saddle geometries for the extended class of thermal SK contours in the field theory parameterized by $\bsig$. We realized that the equality among  boundary time contours is reflected in the bulk as the equality among different choices for slicing up the spacetime, which generated inequivalent SKEW geometries upon analytic continuation. The bulk turned out to be cognizant of the restriction $\Re \bsig \in (0,1)$ in a natural way. We discovered that contrary to the first impression, the CGL prescription does respect the bulk diffeomorphism symmetry, albeit with its own nuances -- analytic variations in $\totr{\csig}$ were genuine gauge symmetries of the SKEW geometry which preserved the boundary SK contour, while  non-analytic variations generated relative translations among the real time segments of the boundary contour. The boundary contour was truly left invariant  by only those bulk diffeomorphisms which vanished at the horizon and continued to the SKEW geometry as an analytic function. In \cref{sec:KScond} we verify a Lorentzian variant of the Kontsevich-Segal  consistency conditions proposed in \cite{Kontsevich:2021dmb,Witten:2021nzp} for SKEW geometries and show that they allow for stable propagation of matter fields.\footnote{A quite similar analytically continued geometry was shown to be stable to D-brane nucleations in \cite{Mahajan:2021maz}, whose authors investigated stability of the double cone geometry which arises in computation of the spectral form factor \cite{Saad:2018bqo}. We expect their results to hold in our case as well. \label{fn:DbraneStable}}

In \cref{sec:RindlerSKEW} we outlined a near horizon version of the construction hoping to understand the universal aspects of CGL like analytic continuation relevant for general spacetime horizons. We clarified why the TFD configuration of SKEW geometries encoded the bulk physics in a manner which is practically equivalent to the usual Kruskal extension. The advantage with the SKEW perspective is that it has no cordinate singularities and hence facilitates the computation of higher point correlation functions via semi-classical Witten-Feynman diagramatics. On the other hand, the absence of Milne wedges in these geometries is a sharp distinction from our traditional perspective on horizons. Moreover, the twisted structure of the embedding of SKEW geometry into the complex null plane we identified, suggests an extension of the SKEW geometry to include the interior is difficult. It will be interesting to study the significance of this observation with respect to holographic constructions like \cite{Gao:2016bin, Maldacena:2017axo, Susskind:2014rva} which utilize or attempt to describe the interior \cite{Papadodimas:2012aq,Papadodimas:2013jku}. 

Another conclusion from the near horizon analysis was that the mechanism behind CGL like prescriptions is essentially the equality between near horizon boost symmetry and the modular Hamiltonian of the bulk, and the latter's identification with the boundary modular Hamiltonian via the JLMS relation.  We also identified an evolution operator on the SKEW contour which corresponds to global Kruskal time evolution (in the TFD configuration with $\bsig = \frac{1}{2}$). This operator was shown to receive localized residue contributions from the horizon, much like the area term in the JLMS relation. We hope to gain a cleaner interpretation of this observation in a future work.

Finally, we suggested that any physical observer describing the exterior would describe the background using an ensemble of coordinate systems. This would naturally lead to a description of the SKEW geometry as a classical ensemble with the moduli $\isig$. We interpreted this as implying the consistency of the open EFT paradigm for the physics on the black hole's exterior with the recent understanding of emergent Type II algebras in semi-classical gravity.

We have already highlighted various potential directions to follow up on our analyses in the main text. An immediate direction is studying semi-classical states excited above the black hole background. Following up the discussion in  \cref{sec:ExThermal}, this involves keeping track of particles entering and leaving the SKEW geometry through the past and future horizons. Another interesting  direction is to more rigorously clarify the relation between the SK ensemble states of \cref{sec:ClassSigEns} and the crossed product algebras of \cite{Witten:2021unn}.

Let us comment on some interesting generalizations of our construction. In this work we discussed CGL like analytic continuations of the Schwarzschild solution in gauges which are  static and homogeneous with respect to the boundary.  One could generalize the diffeomorphism $\totr{\csig}$ in \cref{eq:DiffToIn} by relaxing these conditions, leading to more general configurations of the SKEW geometry. In particular this would involve configurations where the horizon cap itself is dynamic and dressed with localized edge modes. It is interesting to understand what feature of the boundary thermal physics is encoded in such backgrounds. Perhaps this describes a less understood sector in the space of hydrodynamic moduli of the CFT. It is tempting to speculate that the physics at play here relates to the local entropy of the boundary fluid.

A close variant of the SKEW saddle can be constructed  by generalizing the analytical continuation in \cref{eq:AlphaDef} via $f(\alpha) = (\alpha - i 0)^{2 n}$, $n \in \mathbb{Z}_{+}$. This has the effect of changing the radial contour in \cref{fig:mockt} to one which winds around the horizon $n$ times. Consequently, every monodromy in the computation that leads to \cref{eq:PFvar} gets amplified by a factor of $n$, whose net effect is given by $\beta \to n \beta$ in \cref{eq:PFvar}.  This construction is a holographic SK variant of the replica trick which generates correlators with insertions of $\left(\rho_{_{\beta}}\right)^{n}$. The corresponding boundary correlators can be interpreted as those generated from $\mathcal{Z}_{n,m}[J_{\sL/\sR}] = \text{tr} \left\{ \left(\rho_{_{\beta}}\right)^{m}\mathsf{U}[J_{\sR}] \left( \rho_{_{\beta}}\right)^{n-m}    \mathsf{U}^{\dag}[J_{\sL}] \right\}$, where $\Re m \in (0,n)$. Note that since the exterior metric itself is not modified, the boundary correlators generated from $\mathcal{Z}_{n,m}$ would still be built from the original response kernels ($K^{\text{in}}$), but with a KMS periodicity of $n\beta$. The non-trivial aspect of such setups would be tracking localized contributions from the horizon which would now be amplified $n$ times. However, we should perhaps not take these saddles seriously as they likely violate the KS conditions. The natural saddle corresponding to the density operator $\left(\rho_{_{\beta}}\right)^{n}$  is given by the original SKEW construction around a geometry with inverse temperature $n \beta$.

Our analysis straightforwardly applies to de Sitter space, in particular with respect to the paradigm outlined in \cite{Loganayagam:2023pfb}. Those authors defined a cosmological analogue of the boundary influence phase which characterizes the open EFT witnessed by an observer in the static patch of the de Sitter space. Our comments pertaining to SK ensemble observables is applicable to their set up as well, and has similar motivations in light of the results of \cite{Chandrasekaran:2022cip}. A related idea is the construction of SKEW like gravitational saddles for cosmological perturbation theory which focuses on in-in observables anchored to the future time slice of the de Sitter geometry \cite{Chen:2017ryl}.

Finally, we would like to point out an interesting connection between SKEW geometries and the double cone geometries of  \cite{Saad:2018bqo} which compute the late time spectral form factor (SFF). Notice that the radial contour of the CGL prescription is essentially same as the radial contour that defines the double cone, where in the latter the geometry is also compactified along the time direction (see also \cref{fn:DbraneStable}). Our discussion on the KMS properties of boundary correlators and various previous studies on the grSK saddle indubitably argue that the bulk dynamics on these contours compute the SK generating functional \cref{eq:ModSkGen} which involves a single trace in the boundary field theory. Further SK ensemble geometries we introduced sum over the relative time shifts between the boundaries, much like in the double cone family of saddles. However, the SFF involves the squared absolute value of the analytically continued partition function, which corresponds to a product of traces in the boundary theory. How is that identical analytical continuations of the bulk spacetime compute two very different boundary observables? A key difference between the two constructions is that \cite{Saad:2018bqo} also sums over bulk diffeomorphisms which correspond to relative fluctuations in the boundary energy, where as all the gauge choices we discussed on the SKEW geometry only redefined the time coordinate. For example, recall that in  \cref{eq:NullCor}, we always have $v^{+}v^{-} = \rho^{2}$, which generalizes to the full geometry as the statement that large diffeomorphisms we consider do not introduce any relative Weyl factor between the asymptotic boundaries. We conjecture that summing over also such  large diffeomorphisms should effectively modify the single trace in the SK computation to the double trace in the SFF computation. See \cite{Winer:2020gdp} for arguments in this direction, who investigated the connection between SFF and fluctuating hydrodynamics. See also \cite{Chen:2023hra} for related discussions. It would be interesting to study the proposal of \cite{Winer:2020gdp} and related corrections to the holographic SFF from the holographic SK point of view.

\paragraph{Acknowledgement: }The author would like to thank Banashree Baishya, Thomas Faulkner, Alok Laddha, Nima Lashkari, R Loganayagam, Raghu Mahajan, Godwin Martin, Shiraz Minwalla, Onkar Parrikar, Mukund Rangamani, Suvrat Raju, Pratik Rath, Rohit Reghupathy, Ashoke Sen, Shivam Sharma, Omkar Shetye, Jonathan Sorce and Junggi Yoon for related discussions. The author is supported through an appointment to the JRG Program of APCTP by the Science and Technology Promotion Fund and Lottery Fund of the Korea Government, the local governments of Gyeongsangbuk-do Province and Pohang City, and the National Research Foundation of Korea (NRF) grant (2022R1A2C1003182) funded by the MSIT, Korea Government.

\appendix

\section{Kontsevich-Segal conditions}  \label{sec:KScond}

Given a classical solution for the metric, there are numerous way to analytically extend it by continuing either the domains of spacetime coordinates or the parameters within the solution to the complex plane. Such spacetimes and semi-classical gravity on them have proven to be very powerful tools in recent years \cite{Hartle:1983ai, Halliwell:1989dy, Louko:1995jw, Saad:2018bqo,Colin-Ellerin:2020mva}. However, given the inumerable possibilities in making such detours via the complex plane, it is important to understand whether, and when, a particular complexified geometry is allowable as a consistent background. A consistency check for such spacetimes has been proposed in \cite{Witten:2021nzp,Kontsevich:2021dmb} called the \textit{Kontsevich-Segal} conditions (KS). In this section we modify these conditions originally stated for Euclidean signature to the case of real time or pseudo-Lorentzian signature  and show that they are satisfied by generic SKEW geometries. Again, SKEW spacetimes are not strictly Lorentzian everywhere, especially on the horizon cap, which is the reason we term it as pseudo-Lorentzian. According to the Euclidean standpoint in \cite{Kontsevich:2021dmb,Witten:2021nzp}, Lorentzian manifolds lie at the boundary of the space of allowable complex metrics. In a similar vein, we will shortly see that the SKEW geometries almost saturate the Lorentzian KS conditions we impose.

KS conditions follow from demanding that analytically continued metric should allow stable propagation of matter fields. Arbitrary $p$-form fields were suggested to be a useful class of test fields for this check \cite{Witten:2021nzp}.\footnote{A $0$-form field is simply a scalar field.} To state these conditions, let's begin by
recalling the complexified metric given in \cref{eq:AlphaMet}, expressed in terms of the real contour parameter $\alpha \in (-1,1)$. The metric is given by $\dd s^{2} = \gorb + r^{2} \dd \vb{x}^{2}$, where
\begin{equation} 
	\begin{split}
		\gorb
		=& - r^{2} f \left( \dd w + \frac{1- \sig }{r^{2}f} r' \dd \alpha \right)   \left(\dd w - \frac{1 + \sig}{r^{2}f } r'  \dd \alpha\right)  \ , \quad r' = 2 \idim \, r^{d+1} (\alpha- i 0) \ ,
	\end{split}
\end{equation}
where $f= (\alpha - i 0)^{2}$ and we have set $r_{h}=1$ in the above and for the remainder of this section.  The advantage of the above parameterization is that the spacetime integration measure on the geometry written in terms of $\alpha$ is real and positive everywhere. We have 
\begin{equation}
	-\det \gorb = (r')^{2} = 4 \idim^{2} r^{2(d+1)}f   \ , \qquad - \tr \gorb = r^{2}f + 4 \idim^{2} (\sig^{2}-1)r^{2d} = :  \mathscr{T} \ .
\end{equation}
For the above geometry, the KS conditions are given by
\begin{equation}\label{eq:KScond0}
	\Im \left(\sqrt{-g} \, g^{i_{1} j_{1}} \ldots g^{i_{p}j_{p}}  F_{i_{1}\ldots i_{p}} F_{j_{1} \ldots j_{p}} \,  \right)   < 0 \ , \quad   0 \le p \le d+1 \ ,
\end{equation}
where $F$ denotes a $p$-form field. The above condition is demanded locally everywhere on the spacetime and is the generalization of KS conditions in \cite{Witten:2021nzp} to Lorentzian signature.\footnote{Evidently, our convention for the path integral is that the Lorentzian action contributes to the measure as $e^{i S_{\text{Lor}}}$.}

We now proceed to simplify the constraints in \cref{eq:KScond0} following \cite{Witten:2021nzp}. 
Assuming that the metric diagonalizes locally as $g = \text{diag}(\lambda_{i} )$, \cref{eq:KScond0} for SKEW geometries translates to conditions
\begin{equation}\label{eq:KScond}
	\Im \left( \sqrt{-g } \prod_{i=0}^{d}  \lambda_{i}^{ a_{i}}  \right)   < 0  \  , \qquad a_{i} \in \{-1,0 \} \ .
\end{equation}
That is, for any choice of the vector $a_{i}$ with components $0 $ or $ -1$, the above condition has to be satisfied.  The conditions in \cref{eq:KScond} can be simplified in terms of phases (arguments) of its constituent factors, as we discuss now.

The eigen values of $\gorb$ are
\begin{equation}
	\lambda_{0} =  - \frac{1}{2}\left(\sqrt{4(r')^{2}+\mathscr{T} ^{2}} + \mathscr{T}  \right) \ , \quad \lambda_{1} = -\frac{(r')^{2}}{\lambda_{0}} \ .
\end{equation}
The volume measure $\sqrt{-g}$ evaluates to
\begin{equation}\label{eq:rgmeasure}
	\sqrt{-g} := (-\lambda_{0} \lambda_{1})^{1 \over 2}  r^{d-1} = r' r^{d-1} \ ,
\end{equation}
where $\lambda_{0}$ and $\lambda_{1}$ denote eigenvalues along local temporal and radial directions respectively and $\lambda_{i>1} = r^{2} $. We have effectively factorized the determinant between the transverse $\mathbb{R}^{d-1}$ and the orbit space directions. Notice that $\Re r'$ is not positive definite on the contour. This is consistent with the notion that while computing the gravitational action, contributions from the segment with $\Re \ctor =0 $   $ (\alpha<0)$ is subtracted while contributions from the complement segment with $\Re \ctor =1$ $(\alpha>0)$ is added. In this regard, the function $r'$ behaves similar to $\sqrt{f}$ on the contour (in fact they are proportional up to an analytic factor). The parameterization in terms of $\alpha$ achieves precisely this. The horizon cap region interpolating the two segments and is also consistently described by \cref{eq:rgmeasure}. 

With the above convention of defining $\sqrt{-g}$, \cref{eq:KScond} simplifies further to a form similar to the Euclidean KS condition discussed in \cite{Witten:2021nzp}, namely
\begin{equation}\label{eq:KScond1}
	\Im \left( e^{i\eta} \, \prod_{i=0}^{d} \lambda_{i}^{\frac{1}{2} b_{i}} \right) < 0 \ ,  \quad b_{i} = \pm 1 \ , \qquad \eta := \arg \left(  \frac{r'}{\lambda_{0}^{1 \over 2}  \lambda_{1}^{1 \over 2}}\right) \ .
\end{equation}
In order to simplify \cref{eq:KScond1}, we need to closely inspect phases of $\lambda_{i}$ on the contour.  It is convenient to  define the phases such that $\log z \in [- \pi, \pi )$, where $z$ stands for one of the eigenvalues. In other words, we align the branch cut of $\log z$ along the negative real axis.  In order to correctly determine the $\log$ of various quantities featuring in \cref{eq:KScond1}, we need to track them close to the turning point given by $\alpha = 0$. Provided we work with a $\sig$ profile which behaves smoothly on the horizon cap, the near horizon behaviour of $\lambda_{0}$ falls into distinct cases based on whether $\sig^{2}<1$ , $\sig^{2}=1$ or $\sig^{2}>1$ at the horizon.

\begin{figure}[h!] 
	\begin{center}
		\begin{tikzpicture}[scale=0.6]
			\node[draw,color=purple] at (-4,2) {$\scriptstyle{z = \sqrt{f}=\alpha - i0}$};
			\draw[thin, color=teal!50,  ->] (0, -2) -- (0, 2);
			\draw[thin, color=teal!50,  ->] (0,0) -- (4,0);
			\draw[thin, color=teal!50, fill=teal!50] (2.5,0) circle (0.3 ex) node [above] {$\scriptstyle{1}$};
			\draw[thin, color=teal!50, fill=teal!50] (-2.5,0) circle (0.3 ex) node [above] {$\scriptstyle{-1}$};
			\draw[thin, snake light, color=teal!70,  -] (-4,0) -- (0,0);
			\draw[thin, color=teal!50, fill=teal!50] (0,0) circle (0.3 ex);
			\draw[thin, color=purple, ->-] (-2.5,-0.25) -- (-0.25,-0.25);
			\draw[thin, color=purple, ->-] (0.25,-0.25) -- (2.5,-0.25);
			\draw[thin,color=purple, -] (-0.25,-0.25) arc (180:360:0.25);
			\draw[thin, color=purple, fill] (-2.5,-0.25) circle (0.3 ex) node [below] {$\scriptstyle{\ctor = 0}$};
			\draw[thin, color=purple, fill] (2.5,-0.25) circle (0.3 ex) node [below] {$\scriptstyle{\ctor = 1}$};
		\end{tikzpicture}
		\hskip 35 pt
		\begin{tikzpicture}[scale=0.6]
			\node[draw,color=purple] at (-4,2) {$\scriptstyle{z=\lambda_{0}}$};
			\draw[thin, color=teal!50,  ->] (0, -2) -- (0, 2);
			\draw[thin, color=teal!50,  ->] (0,0) -- (4,0);
			\draw[thin, snake light, color=teal!70,  -] (-4,0) -- (0,0);
			\draw[thin, color=teal!50, fill=teal!50] (0,0) circle (0.3 ex);
			\draw[thin, color=purple, ->-] (-4,-0.25) -- (-0.25,-0.25);
			\draw[thin, color=purple, ->-] (-0.25,0.25) -- (-4,0.25);
			\draw[thin,color=purple, -] (-0.25,-0.25) arc (225:495:0.353);
			\draw[thin, color=purple, fill] (-4,-0.25) circle (0.3 ex) node [below] {$\scriptstyle{\ctor=0}$};
			\draw[thin, color=purple, fill] (-4,0.25) circle (0.3 ex) node [above] {$\scriptstyle{\ctor = 1}$};
		\end{tikzpicture}
		\vskip 10 pt
		\begin{tikzpicture}[scale=0.6]
			\node[draw,color=purple] at (-4,2) {$\scriptstyle{z=\sqrt{\lambda_{0}}}$};
			\draw[thin, color=teal!50,  ->] (0, -2) -- (0, 2);
			\draw[thin, color=teal!50,  ->] (0,0) -- (4,0);
			\draw[thin, snake light, color=teal!70,  -] (-4,0) -- (0,0);
			\draw[thin, color=teal!50, fill=teal!50] (0,0) circle (0.3 ex);
			\draw[thin, color=purple, ->-] (0.25,-2) -- (0.25,-0.25);
			\draw[thin, color=purple, ->-] (0.25,0.25) -- (0.25,2);
			\draw[thin,color=purple, -] (0.25,-0.256) arc (315:405:0.3535);
			\draw[thin, color=purple, fill] (0.25,-2) circle (0.3 ex) node [right] {$\scriptstyle{\ctor=0}$};
			\draw[thin, color=purple, fill] (0.25,2) circle (0.3 ex) node [right] {$\scriptstyle{\ctor= 1}$};
		\end{tikzpicture}
		\hskip 35 pt
		\begin{tikzpicture}[scale=0.6]
			%	\draw[thin, color=teal!50,  ->] (-4,2) -- (-4,2.7) ;
			%	\draw[thin, color=teal!50,  ->] (-4,2) -- (-3.3,2);
			%	\draw[very thin, color=teal!50] (-4,2.4) circle (0.001 ex) node [right] {$\scriptstyle{\sqrt{\lambda_{0}}}$};
			\node[draw,color=purple] at (-4,2) {$\scriptstyle{z=\lambda_{1}}$};
			\draw[thin, color=teal!50,  ->] (0, -2) -- (0, 2);
			\draw[thin, color=teal!50,  ->] (0,0) -- (4,0);
			\draw[thin, snake light, color=teal!70,  -] (-4,0) -- (0,0);
			\draw[thin, color=teal!50, fill=teal!50] (0,0) circle (0.3 ex);
			\draw[thin, color=purple, ->-] (0.9,-0.25) -- (4,-0.25);
			\draw[thin, color=purple, ->-] (4,0.25) -- (0.9,0.25);
			\draw[thin,color=purple, -] (0.9,0.25) arc (90:270:0.25);
			\draw[thin, color=purple, fill] (4,0.25) circle (0.3 ex) node [above] {$ \qquad \scriptstyle{\ctor=0 \, }$};
			\draw[thin, color=purple, fill]  (4,-0.25) circle (0.3 ex) node [below] {$ \qquad  \scriptstyle{\ctor = 1}$};
		\end{tikzpicture}
		\caption{Trajectories of $z = \sqrt{f},\lambda_{0}, \sqrt{\lambda_{0}}, \lambda_{1}$ on the complex plane as one moves along the SKEW contour with $\sig^{2}<1$ are shown. The branch cut in $\log z$ is aligned along the negative real axis and we have $\log z \in [- \pi, \pi) $ on the principal sheet. Alternatively this is the image of the SKEW contour on the corresponding $z$ plane. The location of asymptotic cut off surfaces $\ctor=0,1$ are marked. The $\lambda_{1}$  locus wraps around the branch point at $\infty$.}
		\label{fig:EigenTraj}
	\end{center}
\end{figure}
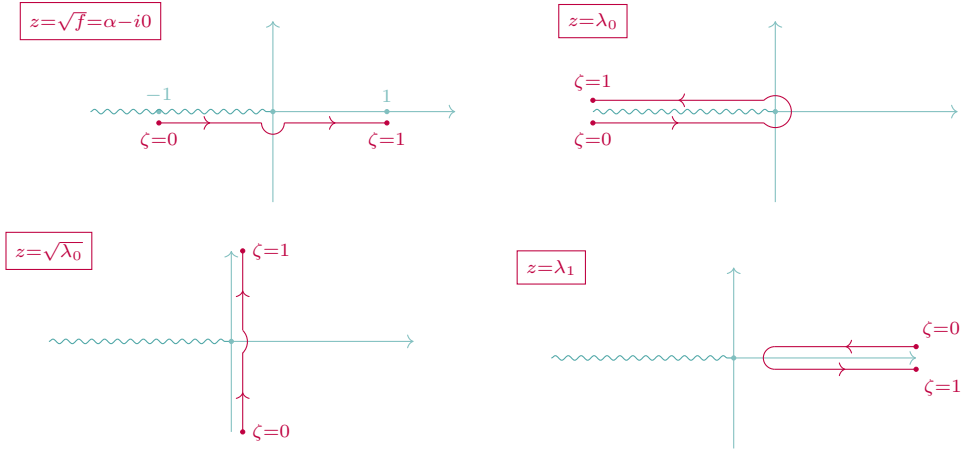

\subsection{Case I: $\sig^{2}<1$ at the horizon} 

In this case, for $|\alpha| \to 1$ (alternatively for  large $r$), we find
\begin{equation}
	\mathscr{T} \sim 4 \idim^{2}(\sig^{2}-1) r^{2d} \ , \quad \lambda_{0} \sim \frac{r^{2}}{\sig^{2}-1} \ , \quad \lambda_{1} \sim - \mathscr{T} \ .
\end{equation}
Near the horizon, $\alpha \to 0 $, and we get
\begin{equation}
	\mathscr{T} \to 4 \idim^{2} (\sig^{2}-1) < 0  \quad  \Rightarrow \quad \lambda_{0}| _{\alpha \to 0 } \to  \frac{f}{\sig^{2}-1} \ , \quad \lambda_{1}|_{\alpha \to 0} \to 4 \idim^{2} (1- \sig^{2}) \ .
\end{equation}
In particular, since $f|_{\alpha=0} < 0 $, we have $\lambda_{0}|_{\alpha = 0} > 0 $. Therefore, as $\alpha$ is varied, $\lambda_{0}$ starts of from $-\infty$ below the negative real axis and winds around the branch cut, returning to $-\infty$ above the cut. On the other hand $\lambda_{1}$ starts off from $\infty$, slightly above the positive real axis, approaches the minimum value at $\alpha=0$, and bounces back to $\infty$ slightly below the real axis . The loci of $z=\sqrt{f}, \lambda_{0}, \sqrt{\lambda_{0}}$ and $\lambda_{1}$ as one traverses the SKEW contour is shown in \cref{fig:EigenTraj}. We choose the $\log z$ branch cut along the negative real axis and therefore have $\log z \in [-\pi, \pi )$ on the principal sheet.  The phases of various quantities of interest behave as follows:
\begin{equation}
	\arg r' \in (-\pi , 0) \ , \quad \arg \lambda_{0} \in (-\pi, \pi) \ , \quad \arg \lambda_{i > 0} \in (-\delta, \delta) \ ,
\end{equation}
where $\delta$ is a small real value which can be made arbitrarily small on the contour. Evaluating $\eta$ separately for the two segments of the SKEW contour, we find they match. In expression, 
\begin{equation}
	\begin{split}
		\alpha &< 0 :   \qquad  \eta = (-\pi + 0 )_{{r'}} - \left(- \frac{\pi}{2}+ 0 \right)_{{\sqrt{\lambda_{0}}} } - (0)_{_{\sqrt{\lambda_{1}}}} \sim - \frac{\pi}{2} \ , \\
		\alpha &> 0 :   \qquad  \eta = (-0 )_{{r'}} - \left( \frac{\pi}{2} - 0 \right)_{{\sqrt{\lambda_{0}}} } - (-0)_{_{\sqrt{\lambda_{1}}}} \sim - \frac{\pi}{2}  \ . \\
	\end{split}
\end{equation}
To clarify our notation, the subscript denotes which factor contributes the term and $0$ denotes a infinitesimal positive quantity. In the final expression for $\eta$, we have suppressed the infinitesimal part and its effect is considered to simply make stricter the inequalities that follow. Since $r' \sim (\alpha - i 0)$ near the horizon, the phases of $r'$ and $\sqrt{\lambda_{0}}$ continue to cancel each other exactly as they wind around branch point, keeping $\eta $ constant on the contour.

The condition in \cref{eq:KScond1} now reduces to 
\begin{equation}\label{eq:KSsafe}
	-\pi < \arg \left( e^{-i \frac{\pi}{2}} \, \prod_{i=0}^{d} \lambda_{i}^{\frac{1}{2} b_{i}} \right) < 0 \quad  \Rightarrow \quad     \qquad 0   <  \sum_{i=0}^{d}  | \arg \lambda_{i} | < \pi \ ,
\end{equation}
which is same as the condition proposed in \cite{Witten:2021nzp}. Since $\arg \lambda_{i>0} \approx 0$, we need to include the contribution of only $\lambda_{0}$. It is easily deduced from \cref{fig:EigenTraj} that $| \arg \lambda_{0} | < \pi $ on the contour, showing that for $\sig^{2}<1$, the corresponding SKEW geometry satisfies KS conditions.

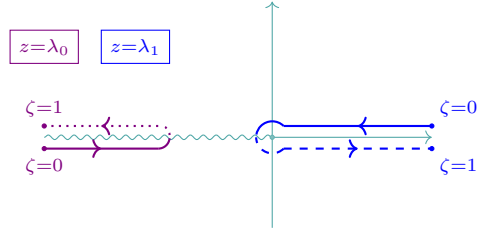
\begin{figure}[h!]
	\begin{center}
		\begin{tikzpicture}[scale=0.6]
			\node[draw,color=violet] at (-5,2) {$\scriptstyle{z=\lambda_{0}}$};
			\node[draw,color=blue] at (-3,2) {$\scriptstyle{z=\lambda_{1}}$};
			%	\node[draw,color=olive] at (-1,2) {$\scriptstyle{z=\sqrt{\lambda_{0}}}$};
			\draw[thin, color=teal!70,  ->] (0, -2) -- (0, 3);
			\draw[thin, color=teal!70,  ->] (0,0) -- (3.5,0);
			\draw[thin, snake light, color=teal!70,  -] (-5,0) -- (0,0);
			\draw[thin, color=teal!60, fill=teal!50] (0,0) circle (0.3 ex);
			%\draw[thin, color=teal!90, fill=teal!90] (2.5,0) circle (0.3 ex);
			%\draw[thin, color=teal!60, fill=teal!60] (-2.5,0) circle (0.3 ex);
			%\draw[thin, color=teal!90, fill=teal!90] (0,-1.8) circle (0.3 ex);
			%
			\draw[thick, color=violet, ->-] (-5,-0.25) -- (-2.5,-0.25);
			\draw[thick, color=violet, dotted, ->-] (-2.5,0.25) -- (-5,0.25);
			\draw[thick,color=violet, -] (-2.5,-0.25) arc (270:355:0.25);
			\draw[thick,color=violet, dotted, -] (-2.25,0.05) arc (5:80:0.25);
			\draw[thin, color=violet, fill] (-5,-0.25) circle (0.3 ex) node [below] {$\scriptstyle{\ctor = 0}$};
			\draw[thin, color=violet, fill] (-5,0.25) circle (0.3 ex) node [above] {$\scriptstyle{\ctor = 1}$};
			
			\draw[thick, color=blue, ->-] (3.5,0.25) -- (0.25,0.25);
			\draw[thick, dashed,  color=blue, ->-] (0.25,-0.25) -- (3.5,-0.25);
			\draw[thick,color=blue, -] (0.25,0.25) arc (45:170:0.353);
			\draw[thick, dashed, color=blue, -] (0.25,-0.25) arc (315:180:0.353);
			\draw[thin, color=blue, fill] (3.5,0.25) circle (0.3 ex) node [above] {$\scriptstyle{\qquad \ctor = 0}$};
			\draw[thin, color=blue, fill] (3.5,-0.25) circle (0.3 ex) node [below] {$\scriptstyle{ \qquad \ctor = 1}$};
			
			%	\draw[thin, color=olive, ->-] (0.25,-4) -- (0.25,-2);
			%	\draw[thin, color=olive, ->-] (-0.25,-2) -- (-0.25,-4);
			%	\draw[thin,color=olive, -] (0.25,-2) arc (315:585:0.353);
			%	\draw[thin, color=olive, fill] (0.25,-4) circle (0.3 ex) node [right] {$\scriptstyle{\ctor = 0}$};
			%	\draw[thin, color=olive, fill] (-0.25,-4) circle (0.3 ex) node [left] {$\scriptstyle{\ctor = 1}$};
		\end{tikzpicture}
		\caption{Trajectories of $z = \lambda_{0,1}$ on the complex plane as one moves along the SKEW contour with $\sig^{2}=1$ are shown. The branch cut in $\log z$ is aligned along the negative real axis and we have $\log z \in (- \pi, \pi]$  on the principal sheet. The segments of $\lambda_{0,1}$ loci outside the principal sheet are indicated as dotted and dashed lines respectively. }
		\label{fig:EigenTrajN}
	\end{center}
\end{figure}

\subsection{Case II: $\sig^{2}=1$ at the horizon}

In this case, we have $\mathscr{T}=r^{2} f$. The near horizon behaviour of $\Lambda_{0}$ is   very sensitive to the exact location of the contour and we find, approximately,
\begin{equation}
	\alpha \to  0 :  \quad 	\lambda_{0} \sim - \text{sgn}(\alpha) \,  r'   \  ,
\end{equation}
which is highly singular. This makes it hard to track the locus of eigenvalues near the horizon. It seems likely that, a more careful examination would prove this case to violate the KS consistency conditions. This conclusion is influenced by the fact that, for $\sig^{2}=1$ the radial direction $\del_r$ is null, where as for $\sig^{2}<1$, $\del_r$ described a space like vector. From the perspective of boundary SK contour as well, the corresponding limits with $\bsig = 0,1$ are ill defined as they describe an unregulated version of the SK time contour where the time ordered and anti time ordered branches of evolution overlap.

\subsection{Case III: $\sig^{2}>1$ at the horizon}

In this case, for $|\alpha| \to 1$ and correspondingly large $r$, we have
\begin{equation}
	\mathscr{T} \sim 4 \idim^{2} (\sig^{2}-1) r^{2d} \ , \quad \lambda_{0} \sim - \mathscr{T}\ ,\quad \lambda_{1} \sim r^{2} \ .
\end{equation}
At the turning point $\alpha \to 0$, we have the limits
\begin{equation}
	\mathscr{T} \to 4 \idim^{2}(\sig^{2}-1) >0  \quad \Rightarrow \quad \lambda_{0} \to  - \mathscr{T} \ , \quad \lambda_{1 } \to \frac{f}{\sig^{2}-1} < 0 \ ,
\end{equation}
since $f|_{\alpha=0}<0$. Tracking the loci of $\lambda_{0,1}$ on the complex plane, we find an important deviation from the case $\sig^{2}<1$.  As $\alpha$ is varied from $-1$ to $1$, $\lambda_{0}$ starts from $-\infty$ below the negative real axis, and intersects the logarithmic branch cut when $\alpha =0$, returning to $-\infty$ above the negative real axis. Therefore for $\alpha>0$, $\lambda_{0}$ lies in the second sheet with respect to the $\log z$ branch cut. $\lambda_{1}$ has a similar trajectory. It start from $\infty$ above the real line, and at $\alpha=0$ cuts the $\log z$ branch cut winding around the branch point  at $0$ in counter-clockwise direction, and returns to $\infty$ below the real line. The loci of $z = \lambda_{0,1}$ are depicted in \cref{fig:EigenTrajN}. The phases of the eigenvalues take the ranges
\begin{equation}\label{eq:ArgUnsafe}
	\arg \lambda_{0} \in (-\pi+\delta , -\pi -\delta) \ , \qquad \arg \lambda_{1} \in (0, 2 \pi ) \ , \quad  \lambda_{i>1} \in (-\delta,\delta) \ ,
\end{equation}
where again $\delta$ is a small real which can be tuned to vanish on the contour. Evaluating $\eta$, we again find it to be constant on the contour. In expression,
\begin{equation}
	\begin{split}
		\alpha &< 0 : \qquad \eta = (-\pi+0)_{r'} - \left(-\frac{\pi}{2}+0 \right)_{\sqrt{\lambda_{0}}} - (0)_{\sqrt{\lambda_{1}} } \sim - \frac{\pi}{2} \ , \\
		\alpha &>0 : \qquad \eta = (-0)_{r'} - \left(-\frac{\pi}{2} - 0\right)_{\sqrt{\lambda_{0}}} -(\pi +0)_{\sqrt{\lambda_{1}}}  \sim -\frac{\pi}{2} \ .
	\end{split}
\end{equation}
$\eta$ is invariant on the horizon cap as the monodromy in $r'$ is excatly canceled by that of $\sqrt{\lambda_{1}}$.

KS conditions again take the form \cref{eq:KSsafe}. From the range of arguments given in \cref{eq:ArgUnsafe}, we find that \cref{eq:KSsafe} is violated. The violation particularly happens on the segment with $\alpha > 0$, where $|\arg \lambda_{0}| + |\arg \lambda_{1}| \approx 3 \pi $. Had we decided to define $\arg \lambda_{1}$ with a shift of $-2\pi$, the violation of KS conditions would have seemed to occur on the region with $\alpha>0$. The takeaway is that, however we fix our definition of the $\arg$ function (or equivalently the position of branch cut of $\log z$), KS conditionss will be violated somewhere on the horizon cap of the radial contour.

Therefore we deduce that, SKEW geometries with $\sig^{2} >  1$ are non-allowable as they violate KS conditions. In particular, the grSK geometry where $\sig =1$ is a singular limit.  However, we do not yet know what implication this subtlety bears on usual observables such as boundary correlation functions. To be cautious, we define correlators with $\bsig \to  0, 1$ as limiting cases of correlations with $0<\bsig <1$.

We conclude this section with the following remark. It is natural to ask if one could define a variant of the SKEW contour by reversing the contour direction.\footnote{This can be achieved by reversing the imaginary shift in $\alpha$ and defining $f=(\alpha+i0)^{2}$.} This would simply swap the convention for time ordered and anti time ordered copies of the boundary. However this not desirable due to the following reasons. We may check a simplified version of the KS condition by considering a massive scalar field probing the spacetime. Clearly, in order for the mass contribution $- i \int \sqrt{-g} \, m^{2} \Phi^{2}$ in the action to weigh down the path integral, we need the segment of radial contour with increasing (decreasing) $r$ to have slightly negative (positive) imaginary volume measure. This is evidently satisfied on the SKEW contour as $\sqrt{-g} = r^{d-1}$.  This property is indeed one of the checks included in KS conditions. Second, in quantum field theory path integrals generating time ordered  (anti time ordered) correlations, the time coordinate is defined with an infinitesimal rotation towards the negative (positive) imaginary axis. The SKEW prescription correctly mimics this infinitesimal Wick rotation as the induced boundary metric (up to the Weyl scaling factor of $r^{2}$) is $\dd s^{2} \sim - f \dd t^{2} + \dd \vb{x}^{2}$, with  $ f \to 1 \mp i 0$ on the time ordered and anti time ordered boundaries respectively.

% Bibliography

\bibliographystyle{JHEP}
\bibliography{BiblioGTC}

\end{document}